\documentclass[preprint2]{aastex}
\def\a{{\bf a}}
\def\af{a_{\rm f}}
\def\B{{\bf B}}
\def\amax{a_{\rm max}}
\def\b{{\bf b}}
\def\Edotdis{\dot{E}_{\rm k}^{\rm diss}}
\def\Edotin{\dot{E}_{\rm k}^{\rm in}}
\def\Ek{E_{\rm k}}
\def\Einps{\dot{e}_{\rm k}}
\def\Es{e_{\rm k}}
\def\epsdot{\dot{\epsilon}_{\rm k}}

\def\sfrOB{\dot{\Sigma}_{\rm OB}}
\def\Deltat{\Delta t_{\rm OB}}
\def\sfrS{\dot{\Sigma}_{\star}}
\def\lf{l_{\rm f}}
\newcommand\mathnew{\mathsurround=0pt}
\def\nstars{n_{\rm s}}
\def\td{t_{\rm d}}
\def\tin{t_{\rm in}}
\def\Sf{S_{\rm f}}
\def\Sigmag{\Sigma_{\rm g}}
\def\simov#1#2{\lower .5pt\vbox{\baselineskip0pt
    \lineskip-.5pt\ialign{$\mathnew#1\hfil##\hfil$\crcr#2\crcr\sim\crcr}}}  

\def\u{{\bf u}}
\def\uf{u_{\rm f}}
\def\urms{u_{\rm rms}}

\def\VS{V\'azquez-Semadeni}
\def\x{{\bf x}}

\def\kms{{\rm km~s}$^{-1}$}
\def\lesssim{{_ <\atop{^\sim}}}
\def\msun{M_{\odot}}
\def\k{{\bf k}}

\slugcomment{Submitted to ApJ, 8th May, 2000}
\shorttitle{Turbulent dissipation in the ISM}
\shortauthors{Avila-Reese \& V\'azquez-Semadeni}
\begin{document}

\title{Turbulent dissipation in the ISM: the coexistence of
forced and decaying regimes and implications for galaxy formation and
evolution} 
 
\author{Vladimir Avila-Reese$^1$ and Enrique V\'azquez-Semadeni$^2$}
\affil{$^1$Instituto de Astronom\'\i a, UNAM, A.P. 70-264, 04510
M\'exico D. F., M\'exico. e-mail: avila@astroscu.unam.mx} 

\affil{$^2$Instituto de Astronom\'\i a, UNAM, Campus Morelia,
A.P. 3-72 (Xangari), Morelia, Mich.\ 58089, M\'exico. e-mail:
e.vazquez@astrosmo.unam.mx}

\begin{abstract}
We discuss the dissipation of turbulent kinetic energy 
$\Ek$ in the global interstellar medium (ISM) by means of
two-dimensional, MHD, non-isothermal simulations in the presence of
model radiative heating and cooling. We argue that dissipation in two
dimensions is 
representative of that in three dimensions as long as it is dominated
by shocks rather than by a turbulent cascade. Contrary to previous
treatments of dissipation in the ISM, in this
work we consider realistic, stellar-like forcing: energy is injected at
a few isolated sites in space, over relatively small scales, and over
short time 
periods. This leads to the coexistence of forced and decaying regimes
in the same flow, to a net propagation of turbulent kinetic energy
from the injection sites to the decaying regions, and to
different characteristic dissipation rates and times in the forced
sites and in the global flow.

We find that the ISM-like flow dissipates its
turbulent energy rapidly. In simulations with forcing, the input
parameters are the radius $\lf$ of the forcing region,
the total kinetic energy $\Es$ each source deposits into the 
flow, and the rate of formation of those regions, $\sfrOB$. The 
global dissipation time $\td$ depends mainly on $\lf$.
We find that for most of our simulations
$\td$ is well described by a combination of parameters of the forcing
and global parameters of the flow: $\td \approx \urms^2/(\epsdot f)$,
where $\urms$ is the rms velocity dispersion, $\epsdot$ is the 
specific power of each forcing region, and $f$ is the filling factor 
of all these regions. In terms of measurable properties of the ISM, 
$\td \gtrsim \langle \Sigmag \rangle
\urms^2/(\Es \sfrOB)$, where $\langle \Sigmag \rangle$ is the average
gas surface density; for the solar neighborhood, $\td\gtrsim 1.5\times
10^7$ yr. The global dissipation time is consistently smaller than
the crossing time of the largest energy-containing scales, suggesting
that the local 
dissipation time near the sources must be significantly smaller than
what would be estimated from large-scale quantities alone. 

In decaying simulations, we find that the kinetic energy
decreases with time as $\Ek(t)\propto t^{-\alpha}$, where
$\alpha\approx$ 0.8--0.9. This
result can be translated into a decay with distance $\ell$ when
applied to the mixed forced+decaying case, giving $\Ek \propto
\ell^{-2 \alpha/(2-\alpha)}$ at large distances from the sources.

Our results, if applicable in the direction perpendicular to
galactic disks, support models of galaxy evolution in 
which stellar energy injection provides significant support for the 
gas disk thickness, but do not support models in which this 
energy injection is supposed to reheat an intra-halo medium at 
distances of up to 10-20 times the optical galaxy size, as the dissipation 
occurs on distances comparable to the disk height. However, this
conclusion is not definitive until the effects of stratification on
our results are tested.

\end{abstract}


\section{Introduction}

The large-scale star formation (SF) cycle in disk 
galaxies plays a key role in modeling
galaxy formation and evolution. This cycle 
crucially depends on the dissipative properties
of the turbulent interstellar medium (ISM).
In normal, noninteracting disk galaxies, SF is believed to be
statistically stationary, and self-regulated\footnote{Throughout the paper, by
self-regulation we only mean that the SF rate feeds back on itself, 
generally in a negative fashion,
maintaining a statistically stationary value over long time
scales, but of course allowing for significant spatial and temporal
fluctuations.} by a balance between the energy injection and 
dissipation in the turbulent ISM in the disk (e.g., Scalo \& 
Struck-Marcell 1984; V\'azquez \& Scalo 1989; Dopita 1990; 
Firmani \& Tutukov 1992, 1994; Dopita \& Ryder 1994; Wang \& Silk 
1994; Firmani, Hern\'andez, \& Gallagher 1996; Avila-Reese 1998; 
Struck \& Smith 1999; Avila-Reese \& Firmani 2000). A self-regulating 
SF mechanism has also been used in semi-analytical models of galaxy 
formation (e.g., White \& Frenk 1991; Kauffmann, White, \& 
Guiderdoni 1993; Cole et al.\ 1994; Somerville \& Primack 1999; 
van den Bosch 1999), but in this case it has been
applied to a hypotethic {\it intrahalo} medium in hydrostatic
equilibrium with its own gravitational potential plus that of the
cosmological dark matter halo, which
extends up to approximately 15-20 times the optical radius of the
galaxy. In these models the disk ISM is virtually ignored and the 
feedback from stars is assumed to efficiently reheat and 
drive back the cooled gas into the intrahalo medium, the reheated
gas fraction being a strong function of the halo mass.
Unfortunately, in most galaxy formation and evolution
models, the detailed thermo-hydrodynamics of 
the ISM has not been treated explicitly. Therefore, several 
assumptions and approximations have had to be made about the SF feedback and 
the dissipative properties of the ISM.

Stars ---mainly massive OB stars and the SN explosions they 
produce--- are sources of thermal and turbulent kinetic energy in
the ISM. Results from  
thermo-hydrodynamical simulations in different contexts 
have shown that the {\it kinetic} energy $\Ek$ injected by SNe 
and massive stars to the ISM deeply affects the dynamics 
of the gas in disk galaxies causing it to be highly
turbulent (e.g., Bania \& Lyon 1980; Chiang \& Prendergast 1985;
Chiang \& Bregman 1988; Navarro \& White 1993; Mihos \& Hernquist 
1994,1996; Rosen \& Bregman 1995; \VS, Passot \& Pouquet 1995; 
Friedly \& Benz 1995; Passot, \VS\ \& Pouquet 1995; Gerritsen 
1997; Avillez 1999; Korpi et al. 1999; see also the 
reviews by Scalo 1987 and by \VS\ et al.\ 2000). Cloud formation and
disruption, the structure and dynamics of the vertical gaseous disk
(in particular its scale height), 
fountains and chimneys, and even a huge hot gas corona (the
``intra-halo medium'') in 
virial equilibrium (including the potential of a huge cosmological 
dark matter halo), might be some of 
the phenomena and processes related to the turbulence 
produced by stellar kinetic energy sources.
A key ingredient in all these phenomena and processes is  
the {\it kinetic energy dissipation rate} in the 
turbulent ISM, since it determines the effectiveness of large-scale SF
feedback on galaxy formation and evolution. 

Dissipation in turbulent incompressible fluids is mainly
controlled by a turbulent cascade from
large to small scales, since it is in the latter where
kinetic energy is dissipated. Several analytical estimates 
and numerical simulations have shown that incompressible MHD
turbulence decays as $t^{-\alpha}$ with $0.7\lesssim\alpha \lesssim 1.0$ 
(Biskamp 1994; Hossain et al. 1995; Galtier, Politano, \& Pouquet
1997). 
In the case of the large-scale ISM, turbulence involves supersonic 
MHD compressible non-isothermal flows, which most likely are 
dominated by shocks. High-resolution 
3D simulations have been used recently to investigate 
dissipation in compressible MHD isothermal flows with parameters 
corresponding to Galactic molecular clouds (Mac Low et al.\ 1998;
Stone, Ostriker \& Gammie 1998; Padoan \& Nordlund 1999; Mac 
Low 1999a). For the decaying regime, Mac Low et al.\ (1998) and 
Stone et al.\ (1998) have found $\Ek$ to decay as $t^{-\alpha}$, 
with $\alpha \sim0.8-1.0$. For the forced regime, Stone et al. 
(1998) and Mac Low (1999a) concluded that the characteristic 
turbulent dissipation time $t_{d}$ is of the order of or smaller 
than the crossing time for the driving (or {\it forcing}) length 
at the rms turbulent velocity $\upsilon_{\rm rms}$, even in 
the presence of strong magnetic fields. In those 
simulations, turbulence has been driven in Fourier space, 
with a fixed kinetic energy injection rate, $\dot{E}_k^{\rm in}$,
generating random large-scale velocity fluctuations with
a constant $\urms$. 
As a result, kinetic energy is injected ubiquitously (i.e., 
everywhere in space), and typically at large spatial scales,
although Mac Low (1999a) also considered intermediate injection 
scales.

The way in which $\Ek$ is injected into the ISM certainly differs
from the prescription used in the studies mentioned above. The scales at
which kinetic energy injection occurs (the regions directly heated or
accelerated by the stellar activity, of sizes of several pc to a few
hundreds of pc), are small compared to the scales of interest in the
large-scale ISM 
(up to a few kiloparsecs).\footnote{Strictly speaking, since such forcing is
confined to a limited range of (small) spatial scales, in
Fourier space it contains components of all spatial frequencies. In
reference to the stellar driving we consider, in
this paper the ``scale of the forcing'' will always
refer to its characteristic size in physical
space.} Besides, the sources are of short
duration and located at discrete, generally isolated sites. 
Moreover, the global energy input rate is not expected to be constant
in general. This situation implies that in the ISM both forced- and 
decaying-turbulence regimes should coexist, the latter being the regions
located far from the energy injection sources. 

In this paper we use two-dimensional (2D) numerical MHD simulations of 
turbulent compressible fluids resembling the ISM of normal 
disk galaxies (V\'azquez-Semadeni et al.\ 1995, 1996; Passot et al.\ 
1995) to explore the ability of the ISM to 
dissipate the turbulent kinetic energy injected by stellar sources.
In order to do this, we
use an ISM-like turbulent kinetic energy 
injection mechanism, based on ``stellar 
winds''  applied randomly in the medium; thus, we
study how the dissipative properties of the flow depend on the 
parameters of the injection sources. 

It should also be noted that the ISM at large scales is far 
from isothermal (see, e.g., Myers 1978), to the extent that this 
non-isothermality has constituted the starting point for multi-phase 
models of the ISM (e.g., Field, Goldsmith \& Habing 1969; Cox \& Smith 
1974; McKee \& Ostriker 1977). Therefore, in this paper we consider
turbulent flows in the presence of parameterized radiative cooling and
background heating (mimicking the heating from a background UV field
and cosmic rays).

In \S\ref{sec:num_meth} we describe the numerical method, the
implementation of the
kinetic energy injection mechanism, and the parameters and initial 
conditions of the simulations. In \S\ref{sec:driven_turb} we
discuss the dissipation time scales for driven (forced) turbulence
simulations using various energy injection regimes, finding that 
most of the injected energy is dissipated locally; we also present
a simple model to calculate $\td$ as a function of 
observable quantities of the ISM. In \S\ref{sec:decay} we explore 
the decay and propagation of the ``residual''
turbulent energy that ``survives'' the strong local dissipation, by 
considering the simulations after turning off the stellar energy 
input sources. In \S\ref{sec:resol_dim} we discuss possible caveats
of our simulations, such as the dimensionality and the low resolution,
and present higher resolution runs which support our conclusions.
Then \S\ref{sec:comparison} compares our results with previous work,
and \S\S\ref{sec:gal_evol} and \ref{sec:gal_form} discuss the
implications of our results for models of galaxy evolution and
formation, respectively. Finally, in \S\ref{sec:concl} we present a
summary and our conclusions.

\section{Numerical method and simulations} \label{sec:num_meth}

\subsection{The model} \label{model}

We use a slight variation of the numerical model presented in a series
of previous papers (V\'azquez-Semadeni et al. 1995, 1996; Passot et al.\ 
1995) and the reader is referred to those papers for details; here
we just sketch the method. The MHD 
equations, including the internal energy conservation equation,
are solved in two-dimensions (2D) in the presence of model terms for
radiative cooling, background heating, and stellar 
winds. While in previous papers self-gravity, the Coriolis force and
large-scale shear were included, here we do not consider them because
of various reasons. First, self-gravity causes
runaway gravitational collapse when the stellar energy injection is
not applied at the densest regions, as we do here (see \S \ref{sec:SF} 
below). Second, the Coriolis force
modifies the dissipation rate at long times in decaying regimes, but
we consider this to be an unrealistic situation (see \S
\ref{sec:decay}). Third, the large-scale shear introduces a non-zero
lower bound to the kinetic energy, which we feel is irrelevant for the 
purposes of the present paper.

The numerical technique is pseudo-spectral
with periodic boundary conditions, and we use a combination of
regular second-order viscosity of the form 
$\mu[\nabla^2 \u + (1/3) \nabla (\nabla \cdot \u)]$, and hyperviscosity 
of the form $\nu \nabla^8 \u$ in the momentum equation. The
hyperviscosity operator is a very steep function of wavenumber $k$,
and, upon adequate choice of the coefficient $\nu$, confines viscous
effects to a much smaller range of scales than the regular second-order
(or lower-than-eigth-order in general) viscosity, allowing a more
extended inertial (self-similar) range in the kinetic energy
spectrum. Our simulations show clear power-law inertial ranges over
more than one decade in wavenumbers (\S \ref{sec:comparison}). It
should be pointed out that finite-difference schemes 
necessarily introduce numerical dissipation which is typically
second-or third-order.
On the other hand, hyperviscosity introduces spurious oscillations 
in the vicinity of strong shocks (Passot \& Pouquet 1988). In order to
``filter'' these out we add a small amount of second-order viscosity,
which, however, can be kept much smaller than what would be necessary
without the hyperviscosity. Note that the use of hyperviscosity should 
not affect the global dissipation rate of the flow, as this rate adjusts itself
to counterbalance the energy injection by the stars, independently of
the specific form of the dissipation operator. What changes from one
operator to another at a given dissipation rate is the dissipation
scale (e.g., Frisch 1995) and the saturated energy content (Stone et
al.\ 1998).

Additionally, the code uses second-order mass diffusion (with coefficient
$\mu_\rho$) with the purpose of
smoothing out shocks even further, as the spectral scheme we use cannot
handle exceedingly steep gradients. We do not expect the mass
diffusion to affect the dissipation rates in the code since its only
purpose is to prevent the formation of excessively small density structures.
The dissipative
coefficients are chosen in such a way that the dissipation operators
are already strongly active a safe margin before the shocks become too 
steep to be resolved.

In the $128^2$ simulations we use $\nu=10^{-11}$, $\mu=0.003$ and
$\mu_\rho=0.06$. We take zero electrical resistivity, and a thermal
diffusivity $\eta=\gamma [\mu+\nu (n/4)^6]$, where $\gamma$ is the
ideal-gas ratio of specific heats, and $n$ is the resolution. All
coefficients are expressed in the code's
non-dimensional units, and are, in the momentum and magnetic flux
equations, essentially inverse Reynolds numbers. 

We have chosen physical units in such a way that the simulations 
resemble the ISM in the plane of the Galaxy near the solar
neighborhood at the 1 kpc scale.
We adopt as normalization values a number density $n_0=1$ cm$^{-3}$,
a temperature $T_0=10^4$ K, a velocity $u_0=11.7$ km s$^{-1}$,
and a length $L_0=1$ kpc. The velocity and length units 
imply a time which (divided by $2\pi$ since in non-dimensional units,
the length of the box is $2\pi$) gives the code time unit, 
$t_0=(L_0/u_0)/2\pi=1.36 \times 10^7$ yr.
The magnetic field is written as $\B=B_0 e_x+\b$, where  
$B_0=1.5 \mu$G represents the uniform azimuthal 
component of the field and $\b$ is a superposed fluctuating component of 
rms amplitude $=5 \mu$G.

\subsection{``Star'' formation and kinetic energy injection} \label{sec:SF}

In previous papers, an ``OB star'' (or, in general, a {\it source})
was turned on at grid point $\x$
whenever the density exceeded a certain (arbitrary) threshold value
and the velocity field had finite negative divergence. A ``star''
consisted of a local source of heat, which caused the nearby gas to
expand, forming warm bubbles resembling HII regions. After some
experimentation, in this paper we
have chosen to use ``winds'' (local outward radial accelerations,
see \VS\ et al.\ 1996) rather than heating, and to place the stars 
randomly with a given probability, rather than at the density peaks. 

The reasons for our choice are that the heating scheme deposited
unspecified amounts of kinetic energy into the flow, the exact amount
depending on the local cooling rate, which in turn depends on the
local density and temperature, rendering the measurement of the
kinetic energy injection rate impossible. Moreover, the
density-directed placement of the sources caused three phenomena
which obstruct the study of how the results depend on the
the initial parameters of the energy injection mechanism. First, the
OB star formation rate (SFR) fluctuated strongly, with alternating 
periods of dormancy and of strong activity (see \VS\ et al.\ 1995). 
This complicates the determination
of the characteristic energy injection time, $\Ek/\Edotin$, where
$\Edotin$ is the global energy injection rate, when the latter goes to
zero. Second, the density-directed SFR produces strong clustering of 
the sources, because generally more than one grid point satisfy the
density criterion within a density peak (a ``cloud''). This causes their
outputs to interfere destructively (since oppositely directed accelerations
acting on the same grid point cancel out) in such a way that the
total energy input of a cluster is smaller than the sum of the inputs
from its constituent stars taken separately. Besides, due to clustering,
large ``supershells'' form; these supershells are energy input sources
which cannot be easily related to the initial injection parameters.
In fact, destructive interference between the sources and ``supershell''
formation still occur in the random-star placement scheme at the 
largest SFRs, but to a much lesser extent. Finally, the density-directed 
scheme involves the presence of an initial negative velocity divergence 
at the forcing sites, making it impossible to determine the final 
velocity attained by the flow around the sources, since the acceleration 
acts on a pre-existing velocity field of fluctuating magnitude.
In summary, the random placement of stars we use here, although
certainly less realistic than the density-directed placement, makes
for a much more controlled energy input, and we adopt it in our forced 
simulations.

\begin{figure}
\plotone{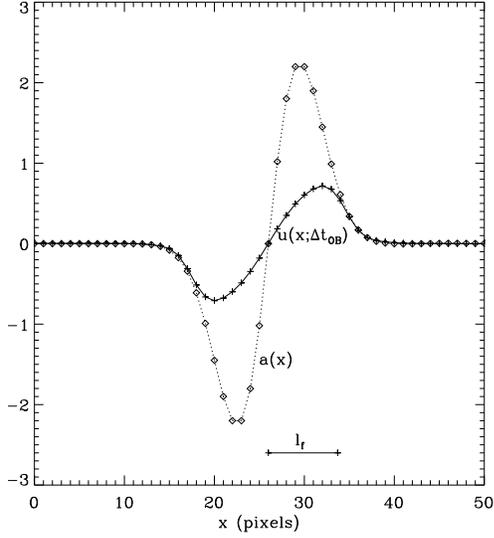}
\caption{One-dimensional profiles of the acceleration $a_x(x)$ ({\it
dotted line}) and the final velocity after applying $a_x(x)$ over
a stellar lifetime ($\Delta t_{\rm OB} =6.8$ Myr) for the parameters
of the fiducial 
simulation labeled run 30. The amplitude $\amax$ is measured from peak
to trough of the $a_x(x)$ curve.}
\label{fig:au_profiles}
\end{figure}

We add a fixed number of sources at each code timestep by
specifying the probability with which a source is placed in each
grid point. However, since the timestep is variable in the code, the
instantaneous number of sources $\nstars$ in the simulation is
not perfectly constant in time. This is why we prefer to use the 
SFR per unit of area, $\sfrOB$, measured directly from the simulations, 
in such a way that at any one time $\nstars = \sfrOB \ \Deltat$, where 
$\Deltat=6.8$ Myr is a typical lifetime of OB stars.
Once a source has been turned on at a given grid
point \x, it stays on for a time $\Deltat$. During this time the gas around
the source receives an acceleration $\a$ directed
radially away from \x, with a magnitude 
\begin{equation}
a(r)=\frac{\amax r}{\sigma} \exp\bigl[-(r^2-\sigma^2)/2\sigma^2\bigr], 
\label{eq:force_profile}
\end{equation}
where $\amax$ is the amplitude of the
acceleration, $r$ is the distance from the position
of the star, and $\sigma$ is a free parameter of the order of a 
few pixels. At a 
resolution of 128 grid points per dimension, 1 pixel $\approx 7.8$ pc. 
The acceleration $\a$ produces an evolving velocity profile $v(r,t)$ 
around the source. We show an example of the acceleration and 
velocity profiles at the end of a stellar lifetime in fig.\ 
\ref{fig:au_profiles} for the fiducial simulation called run 30 
(see Table 1). It is easily seen that $a(r)$ has its maximum
at $r=\sigma$, and that by $r=2.2 \sigma$, it has decreased to $0.2\amax$. 
We choose this as a characteristic radius
of the forcing region, which we denote $\lf$.
It is important to note that $\lf$ is {\it not} the final radius of the
expanding shells that result, which are actually a response of the
flow to the energy injection mechanism, and expand to sizes quite
larger than $\lf$. This is particularly noticeable when stars cluster, 
in which case the shells may reach sizes of a few hundred parsecs,
becoming ``supershells''. 

The characteristic power $\Einps$ and energy $\Es$ injected to
the flow by one source can be roughly estimated as
\begin{eqnarray}
\Einps = \int _{\Sf} \Sigmag \u\cdot\a d^2x \approx \langle 
\Sigmag \rangle _{\rm f} \uf \af \pi \lf^2 = M_{\rm f}\uf\af \nonumber \\
\Es = \Einps \Deltat,
\label{eq:einpsource}
\end{eqnarray}
where $\u$ is the flow velocity vector, 
$\Sf=\pi \lf^2$ is the area of the forcing region, 
$\langle \Sigmag \rangle _{\rm f}$ is its initial average surface 
density, and $\uf$ and $\af$ are characteristic values of the
velocity and acceleration, respectively. Note that, since the
simulations are two-dimensional, it is necessary to specify some
effective disk height $H_{\rm f}$ for the forcing regions in order to 
define their surface density and mass. In the Galaxy, the regions
of highest young cluster density oscillate above and below the plane
by $\sim 50$ pc (Alfaro, Cabrera-Cano, \& Delgado 1991). Therefore, 
we take $H_{\rm f}=100$ pc. Then, $\langle \Sigmag \rangle_{\rm f}
=\rho_ 0 H_f$ with $\rho_ 0=m_{\rm H} \ n_0$, and 
$M_{\rm f}=\Sf\langle \Sigma_{\rm g} \rangle_{\rm f}$; 
we use the simulation mean density $n_0$ as representative 
of the mean density in the forcing regions because
the sources are seeded randomly in the simulation.

The product $\af \uf$ 
may be defined as the characteristic power per unit of mass injected
per forcing region, $\epsdot$. We define 
$\uf\equiv u(\sigma, t=\Deltat) < u_{\rm max}(t=\Deltat)$ and 
$\af$ as the value of $a(r)$ where the area contained under 
the acceleration profile is divided horizontally in two equal 
regions. This roughly happens
at $r=2\sigma$, where $a(r)=0.45\amax$; therefore, we take $\af=0.45\amax$.
These values were decided upon empirically, but we have checked that 
they apply reasonably well to all of our simulations.
Note that $\uf$ is not trivially related to $\af$, $\Deltat$ and $\lf$, 
due to the shape of the acceleration profile and to the fixed time 
over which the acceleration is active. In particular, we have checked 
that the simple formula $v^2\sim 2 \af \lf$ does not apply very accurately.   

Together, the three parameters $\af$, $\lf$ and $\uf$ completely
define the local energy injection process.\footnote{We wish to stress here
that while the former two parameters are controlled by us, the latter
is already a response of the flow to the first two, and has to be {\it 
measured} from the simulations, rather than being user-defined. In
this sense, our model here is still semi-empirical.} However, $\af$
and $\uf$ have meaning mostly within the context of our simulations
only, while $\Einps \propto \uf\af\lf^2 H_{\rm f}$ (or, equivalently, $\Es$,
since we take  $\Deltat=6.8$ Myr=const in all cases) is a more
physically and observationally relevant quantity. Thus, in what
follows, we will consider $\Es$ and $\lf$ as the determining
parameters of the local properties of the injection process.
Additionally, there is a third, global, parameter, namely $\sfrOB$.
In the following 
sections we explore how the dissipative properties of the turbulent 
flow depend on the forcing parameters: $\lf$, $\Es$, and 
$\sfrOB$.

\subsection{The simulations}\label{sec:the_sims}

We have performed a sizeable number of simulations, summarized in Table 1. 
Most of the
simulations have been done at a low resolution of 128 grid points per
dimension, in order to achieve reasonable coverage of parameter space. The
dependence of the dissipation rate on 
resolution is further discussed in \S \ref{sec:resol_dim}, where test
runs at higher resolution ($512^2$) are presented as well, for which
the results obtained from the low-resolution simulations continue 
to hold.

We adopt the simulation labeled run 30 as the ``fiducial'' case.
The set of simulations was initially chosen in such a way 
as to have at least two different values of each of the three parameters 
$\lf, \uf,$ and $\nstars$, for (roughly) constant values of 
the others (see Table 1). However, 
as our study progressed, we realized that the most relevant parameters are 
not exactly those we considered initially and, 
as a result, this philosophy is not clearly reflected in the set of 
derived parameters $\lf$, $\Es$, and $\sfrOB$. In 
any case, the three derived input parameters span a reasonable range
of values and, as will  
be described in \S \ref{sec:driven_turb}, we have constructed a 
simple model which combines them, to which our simulations 
will be seen to conform to within $\lesssim 20$\%.

In the fiducial run, the parameters of the kinetic energy 
injection were chosen to represent a compromise between values typical 
of expanding HII regions and SN remnants at late evolutionary stages
(unfortunately, our code cannot handle very strong shocks). 
For run 30, the power and the total kinetic energy injected by each source 
is $\Einps = 1.35 \times 10^{35}$ erg s$^{-1}$ and $\Es = 2.84 \times 
10^{49}$ erg, respectively (see Table 1). The estimated radius
of a typical HII region before the SN explosion is $\sim 50-60$ pc;
at this time the kinetic energy deposited by the region
--expanding with a velocity of $\sim 10$ \kms\ on average--
into the ISM is $\sim 1-2\times 10^{49}$ erg (e.g., Elmegreen 1991).
The cooled shell swept up by a SN (type II) ejecta has roughly 
the same amount of kinetic energy, although
this energy is injected in a time period much smaller than
in the case of the HII region stage.

\begin{figure}
\plottwo{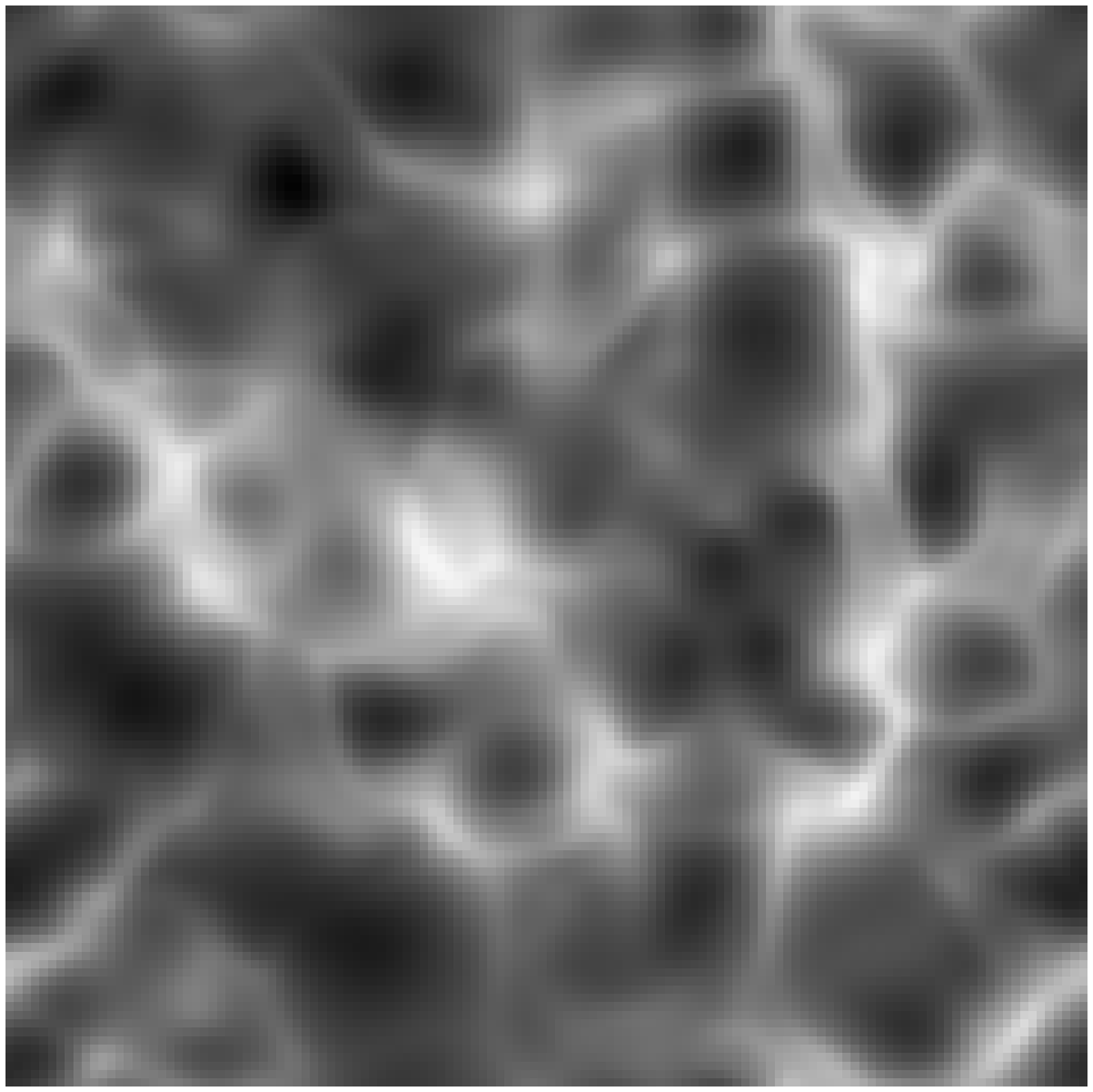}{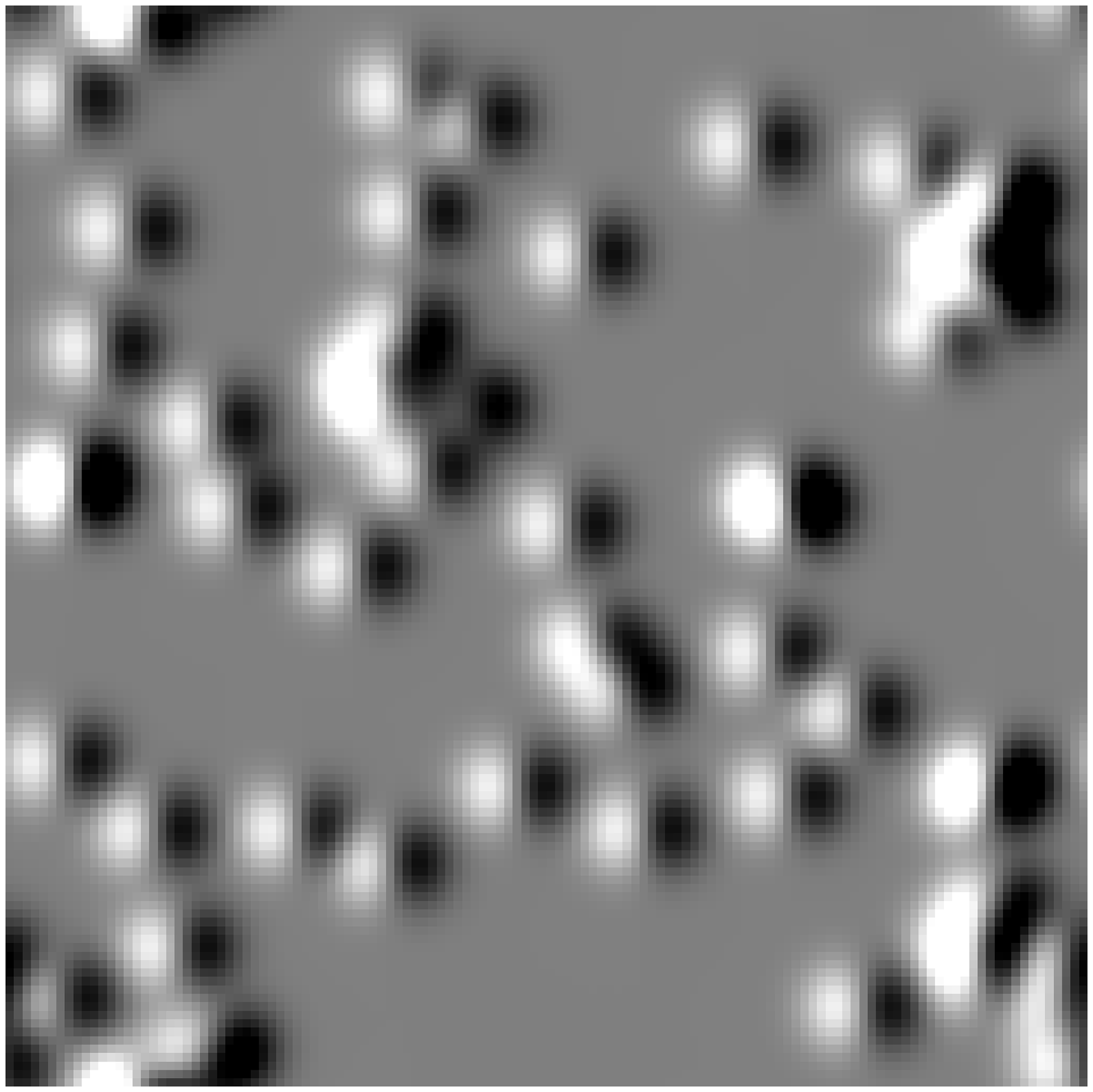}
\caption{Images of the density field ({\it left}) and the
$x$-component of the acceleration for the fiducial run at time $t=272$
Myr. The box size is 1 kpc and the resolution is 128 grid points per
dimension.}
\label{fig:rho_xacc}
\end{figure}

We start the simulations with uniform density and temperature and nearly zero
velocity everywhere ($\urms \sim 0.1$km s$^{-1}$), and begin placing sources
randomly as described in 
\S \ref{sec:SF}. After one source lifetime, the number of sources in
the simulation reaches an approximately stationary number. It is not
precisely constant because of the probability-based algorithm we use
and the variable timestep. The sources generate expanding shells
which interact in a complicated fashion, producing a network of
filaments and distorted 
shells, in a statistically stationary regime. Figure \ref{fig:rho_xacc} shows 
the density and $x$-component of the acceleration at a typical instant
in the evolution of the fiducial simulation, run 30.

We have also performed a simulation with the density-directed scheme
of source placement (run 6) in order to compare our results to this
more realistic simulation. The density threshold used for turning on 
a source was $4 \rho_0$. The input parameters for this run 
were $\lf = 30$ pc and $\uf = \sim 7$ \kms\ (estimated from the final
expansion velocity of the shells).
Finally, in order to study a strictly decaying regime, we 
performed a simulation without SF activity (run 12 in Table 1). This
run used a late evolutionary stage of run 6 as its initial condition.

\section{Dissipation in driven turbulence}\label{sec:driven_turb}

\subsection{The global dissipation rate} \label{sec:global}

\begin{figure}
\plotone{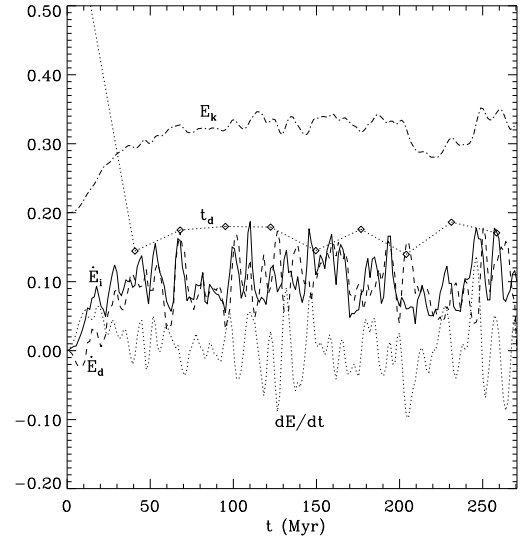}
\caption{Evolution of various quantities for the fiducial run, run 30:
total kinetic 
energy $\Ek$ ({\it dot-dashed line}); total time derivative of the kinetic
energy, $d\Ek/dt$ ({\it dotted line}); kinetic energy injection rate
$\dot{\rm E}_i$ ({\it solid line}); kinetic energy dissipation rate 
$\dot{\rm E}_d$ ({\it dashed line}), as defined by eq.\ (\ref{eq:rates}), 
and the
dissipation time $\td$, defined by eq.\ (\ref{eq:td}) ({\it dotted line
with diamonds}). The latter is shown averaged over periods of 20
consecutive code outputs, due to its sensitivity on $\dot{\rm E}_i$, 
and is given
in units of $10^8$ years. The other quantities are given in 
code units.
}
\label{fig:evol}
\end{figure}

\begin{figure}
\plotone{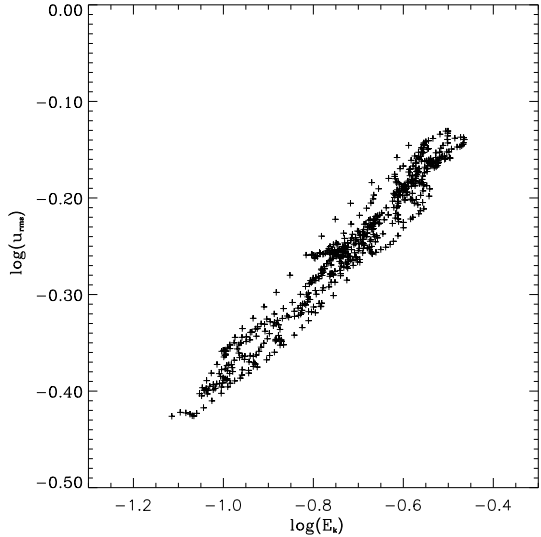}
\caption{Scaling of the rms velocity dispersion $\urms$ vs.\ $\Ek$ for 
run 6. The points define a line with a slope very close to 1/2.}
\label{fig:EvsU}
\end{figure}

For all the forced runs described in \S\ref{sec:the_sims}
we measure several {\it global} quantities of the
flow: the kinetic energy content, $\Ek$, the kinetic energy 
injection rate, $\Edotin$, the total rate of change of $\Ek$, $d\Ek/dt$, 
and the rms velocity dispersion, $\urms$. Figure \ref{fig:evol} shows 
the evolution of these quantities for the fiducial run except $\urms$, 
which we do not plot because, as shown in Fig. \ref{fig:EvsU}, it 
scales very closely as $\Ek^{1/2}$ (see also Mac Low 1999a).
These global quantities are seen to be statistically stationary,
without exceedingly strong fluctuations in time. The injection rate
$\Edotin$ is seen to fluctuate moderately, typically some 50\% around
its mean value. This is not only because the number of sources 
varies slightly in time, but also because of the complex interaction
between the acceleration from the sources and the pre-existing
velocity field (see eq.\ [\ref{eq:force_profile}]).
Similar behaviour is observed in all forced runs with random source
placement. However, a much more irregular behavior is present (not
shown) in the run with density-directed SF, in which the
SFR alternates between zero and strong bursts. Table 2 gives
the average of these quantities between $t=40.8$ and 272 Myr for all runs. 

From the measured values of $\Ek$ and $\Edotin$, we can
define the energy {\it injection} timescale as
\begin{equation}
t_{\rm in}\equiv \Ek/\Edotin.
\label{eq:tin}
\end{equation}
However, the actual quantity 
of interest is $\td$, the kinetic energy {\it dissipation} timescale,
defined as 
\begin{equation} 
\td=\Ek/\Edotdis,
\label{eq:td}
\end{equation}
where $\Edotdis$ is the energy dissipation rate. In a perfectly
stationary regime, $\Edotin$ and $\Edotdis$ coincide, as in the 
well-known Kolmogorov (1941) theory in the incompressible case, and 
in the simulations of Stone et al.\ (1998) and Mac Low (1999a) in the 
compressible case. However, in the more realistic case of a 
regime stationary only in an averaged sense, the two rates need not
coincide. We thus estimate $\Edotdis$ from the equation: 
\begin{equation} 
\frac{d\Ek}{dt}=\Edotin-\Edotdis,
\label{eq:rates}
\end{equation}
where $d\Ek/dt$ can be measured directly in the simulation
from the evolution of $\Ek$. Afterwards, $\td$ can be computed from eq.\
(\ref{eq:td}). The evolution of $\Edotdis$ and $\td$ is also shown in
fig. \ref{fig:evol}. 
The latter quantity is shown averaged over periods of 20 successive
outputs of the code, because of its high sensitivity to fluctuations
in $\Edotin$. The values of $\Edotdis$ and $\td$ averaged
between $t=40.8$ and 272 Myr are given also in Table 2. 

It should also be noted that eq.\ (\ref{eq:rates}) is actually
incomplete, and should include transfer terms from (to) kinetic to
(from) other energy forms, such as magnetic or thermal. Thus, such
transfer terms are effectively included in $\Edotdis$. In particular,
this includes kinetic energy losses due to 
first converting it to those other forms and then
dissipating these by resistivity or radiative cooling. On the other
hand, this implies that $\Edotdis$ may also act as a {\it source} 
of kinetic energy, in which case it may become negative. However, it
is seen that $\Edotdis < 0$ only during a brief period at the beginning 
of the simulation, which, together with the similarity between
$\Edotin$ and $\Edotdis$ discussed below, reassure us that the error
incurred by omitting the transfer terms in eq.\ (\ref{eq:rates}) is
not too severe.

Remarkably, fig. \ref{fig:evol} shows that
$\Edotin$ and $\Edotdis$ are always very close to each other.
Since the energy injection occurs at very localized and scattered
sites, while the dissipation rate is computed over the whole flow,
their highly similar evolutions suggest that {\it most of the dissipation
must occur at or near the injection regions}, and dominates the global 
dissipation rate\footnote{{\it A posteriori}, this is actually not
surprising; in 
stationary forced incompressible regimes, it is well known that 
the dissipation rate is proportional to $\Ek^{3/2}$ 
(see, e.g., Frisch 1995; see also Stone et al.\ 1998; Mac Low 1999a
for the compressible case). Since the kinetic
energy density is greatest at the injection sites, most of the
dissipation must occur there as well. Thus, the dissipation rate is
actually a function of position (through the local kinetic energy
density) and may fluctuate strongly from one position to another, as
is also the case in incompressible turbulence (see, e.g., Frisch 1995).}. 
Only in this case can the global dissipation rate track the 
locally-originated injection so closely. In turn, this
implies that only a small fraction of the energy injected at the
sources ``escapes'' to more distant regions. We refer to this as the
``residual'' turbulence, which coexists in the flow with the forced
regions. We discuss the propagation of turbulent kinetic energy from the
injection to the quiescent sites in \S \ref{sec:decay}. As a
consequence, in the remainder of this paper we refer to the injection
and the dissipation times indistinctly.

Analyzing all the forced runs, we find that $\td$ ($\tin$)
does not significantly depend on two of the energy injection 
parameters, namely the energy input rate of the source, $\Einps$ (or 
just its total energy input $\Es$) and the source formation 
rate, $\sfrOB$ (see \S\ref{sec:SF} for definitions). The relevant 
forcing parameter seems to be the forcing scale, $\lf$. We obtain
the following approximate empirical correlation of $\td$ on the 
input parameters of the forcing:
\begin{equation}
\td\propto \frac{\lf^{0.60}}{(\epsdot \ \sfrOB)^{0.12}}\propto 
\frac{\lf^{0.84}}{(\Es \ \sfrOB)^{0.12}}.
\label{eq:empiric}
\end{equation}

\subsection{Relating $\tin$ to measurable properties of the ISM}
\label{sec:theory}

In order to estimate theoretically $\td (\approx\tin)$ as a function 
of frequently used astronomical quantities of the ISM, we can use 
the definition of $\tin$:
\begin{equation}
\tin\equiv \Ek/\Edotin=
\frac{\int_A \Sigmag u^2 dA'}{\int _A \Sigmag \a\cdot\u dA'}.
\label{eq:tint}
\end{equation}
The integral in the numerator can be estimated as
\begin{equation}
\int _A \Sigmag u^2dA'\approx \langle \Sigmag \rangle \urms^2 A, 
\label{eq:eint}
\end{equation}
where $\langle \Sigmag \rangle$
is the average gas surface density and $A$ is the total area
where the flow is contained. Similarly to what was done in eq.\
(\ref{eq:einpsource}), the integral in the denominator can
be estimated as follows:
\begin{eqnarray}
\int _A \Sigmag \a\cdot\u dA'=\int _{A_{\rm f}} \Sigmag \a\cdot\u dA'\approx
\langle \Sigmag \rangle _{\rm f} \af \uf A_{\rm f} \nonumber \\
=\langle \Sigmag \rangle _{\rm f} \epsdot A_{\rm f},
\label{eq:einpint}
\end{eqnarray}
where the first equality follows because $\a=0$ outside of the forcing
regions, whose total area is $A_{\rm f}$. 
Therefore, from eqs.\ (\ref{eq:tint}), (\ref{eq:eint}), and 
(\ref{eq:einpint}) we obtain
\begin{equation}
\tin\approx \frac{\urms^2}{\epsdot f},
\label{eq:tifin}
\end{equation}
where $f=A_{\rm f}/A$ is the filling factor of forcing regions.
Naively, one could take $f=\pi \lf^2 \nstars / A$. However, one
has to take into account the fact that some of the forcing regions
overlap; the probability of this occurring is larger at larger $\lf$.
Thus, $f\lesssim \pi \lf^2 \nstars / A$ in general, and in practice
we prefer to measure $f$ directly from the simulation as the fraction
of the simulation area in which the stellar acceleration is larger
than 20\% of $\amax$ (see eq.\ [\ref{eq:force_profile}]), corresponding 
to the area within a distance $\lf$ from the centers of the forcing. 
 
Table 2 gives the values of $\urms$ and $f$ 
measured in all the forced runs ($\epsdot$ is an input parameter
given in Table 1), as well as the values of $\tin$ resulting from them. 
These ``theoretical'' injection times are seen to be reasonably close
to those measured directly in the simulations 
according to eq. (\ref{eq:tin}). For
most of the runs, the agreement is better than 20\%, with the largest
deviations being $\sim 45$\% and $30$\%. 

{\it A posteriori}, it is not surprising that eq.\
(\ref{eq:tifin}) gives estimates for $\tin$ consistent
with the values obtained using eq.\ (\ref{eq:tin}), since, after 
all, we have done nothing 
more than estimating the integrals intervening in eq.\ (\ref{eq:tint}) 
in terms of averaged quantities. However, the relevance of this result
is twofold. First, it shows that reasonable estimates can be obtained
using global averaged quantities even when the dissipation is
extremely localized, and second, it provides a means of relating the
dissipation time to frequently used quantities of the ISM. Indeed, using the
expression $f \lesssim \pi \lf^2 \nstars / A$ for the filling 
factor in eq.\ (\ref{eq:einpint}) we obtain 
\begin{eqnarray} 
\int _A \Sigmag \a\cdot\u dA'\lesssim \langle \Sigmag \rangle _{\rm f}
\epsdot\pi\lf^2\nstars= \nonumber \\
M_{\rm f}\epsdot\Delta t_{OB} \sfrOB A=\Es \sfrOB A,
\label{eq:obs}
\end{eqnarray} 
where $M_{\rm f}= \langle \Sigmag \rangle _{\rm f} \pi \lf^2$, 
$\Es=M_{\rm f}\epsdot\Delta t_{OB}$, and $\sfrOB$ has been defined in
\S \ref{sec:SF}. Using eqs.\ (\ref{eq:eint}) and (\ref{eq:obs}) in 
eq.\ (\ref{eq:tint}), one obtains
\begin{equation}
\tin \gtrsim \frac{\langle \Sigmag \rangle \urms^2}{\Es \sfrOB},
\label{eq:obs1}
\end{equation}
which is our desired connection to measurable quantities of the ISM. 
For example, taking $\urms =8$ \kms, $\sfrOB=3\times 10^{-5}$ yr$^{-1}$
 kpc$^{-2}$ 
(Tammann, Loffler \& Schroeder 1994), $\Es=3.5\times 10^{49}$ erg,
and the gas surface density typical for the solar vicinity,
$\Sigmag\approx 12 \msun$pc$^{-2}$, we find that 
$\tin\gtrsim 1.5\times 10^7$ yr. Note that the value of $\Es$ we used
reflects a contribution of roughly $1.5 \times 10^{49}$ ergs per HII
region plus $\sim 2 \times 10^{49}$ ergs of {\it mechanical} energy
per SN event. The resulting value of $\tin$ is in rough agreement with
our fiducial simulation. 

It is worthwhile to note that $\tin$ depends on 
$\Sigmag$ and $\sfrOB$, and since both of these vary with
Galactocentric distance, one expects $\tin$ (and consequently the
dissipation time) to exhibit such a dependence. This question will be
discussed in further detail in section \S\ref{sec:gal_evol}. 

One more point should be remarked regarding eq.\ (\ref{eq:tifin}) (or, 
equivalently, relation [\ref{eq:obs1}]), namely that
the numerical simulations clearly show that the dissipation time 
increases with the size of the forcing regions (see eq.\ 
[\ref{eq:empiric}]), even though relation (\ref{eq:obs1}) has no 
explicit dependence with $\lf$. This is equivalent to asking the 
question: how does the empiric expression for $\tin$ (eq.\ 
[\ref{eq:empiric}]) connect with relation (\ref{eq:obs1}),
which was derived directly from the definition of $\tin$? 
These two relations imply that $\urms$ must scale approximately as
$\lf^{0.42} (\Es \sfrOB)^{0.44}$. 
Recalling that in relation (\ref{eq:obs1}) the inequality accounts for
overlapping of the regions, one expects that an extra factor 
depending on $\lf$ must be included in the RHS of relation (\ref{eq:obs1}) 
in order for it to become an equality.
In the simulations we empirically find that roughly 
$\urms \propto \lf^{0.40} (\Es \sfrOB)^{0.42}$. A full
MHD theory of the problem should account for this dependence.
  
\subsection{A more realistic simulation}

In the runs presented above, we have not included self-gravity nor the 
Coriolis force, and the energy input sources were
placed randomly with a fixed probability, in order to have good
control of the forcing parameters, and to uncover the dependences 
presented in $\S\S$ \ref{sec:global}\ and \ref{sec:theory}, at 
the expense of some realism. Somewhat more realistic simulations 
should include those additional agents, and involve a density-directed
placement of the sources. In particular, the latter scheme causes 
the energy injection rate to fluctuate strongly 
and the input sources to be highly clustered; large supershells
form and significant cancellation of the energy input from
neighboring sources occurs (see \S\ref{sec:SF}). It is thus
interesting to see if our results still hold in this case.

\begin{figure}
\plottwo{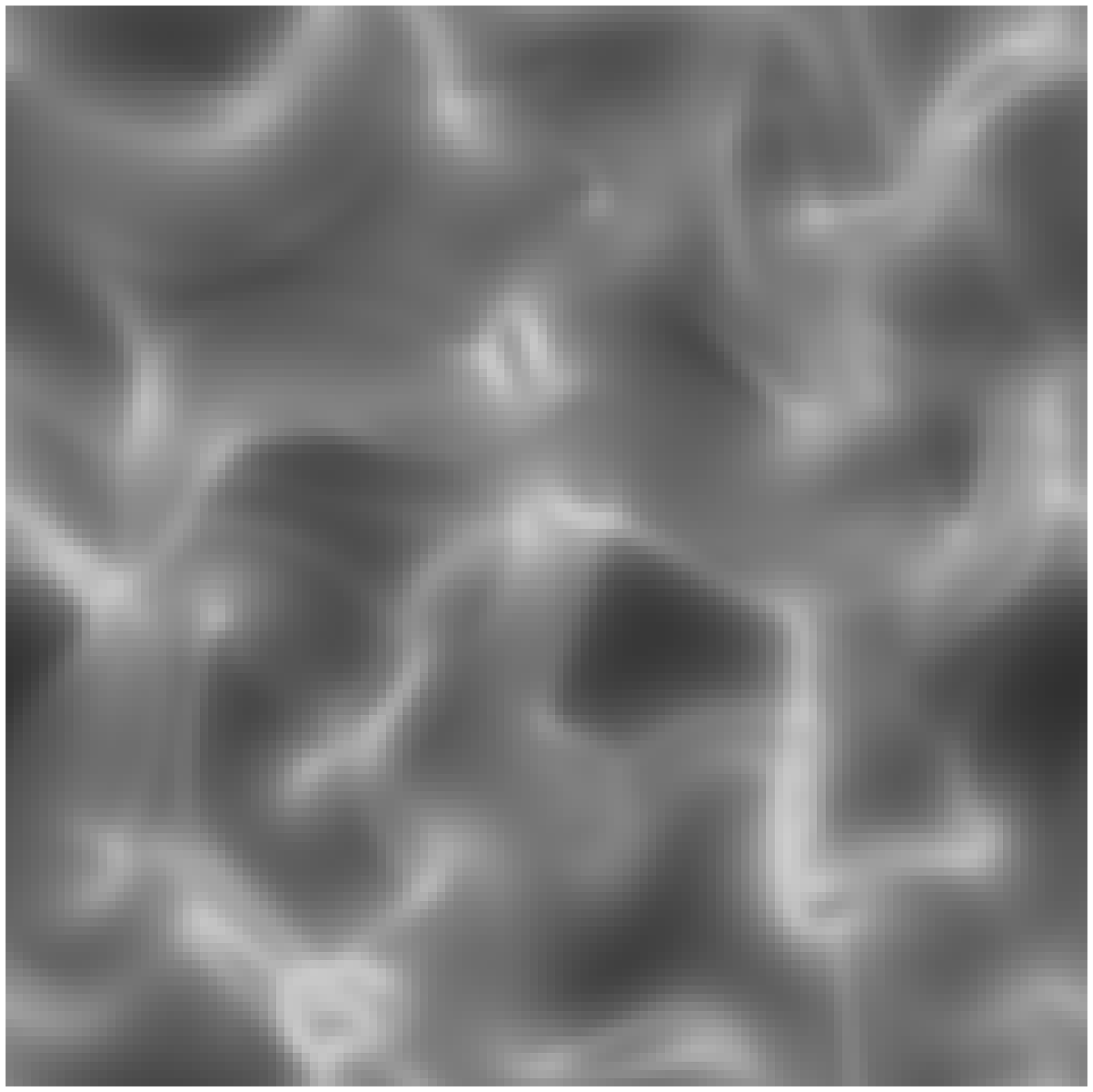}{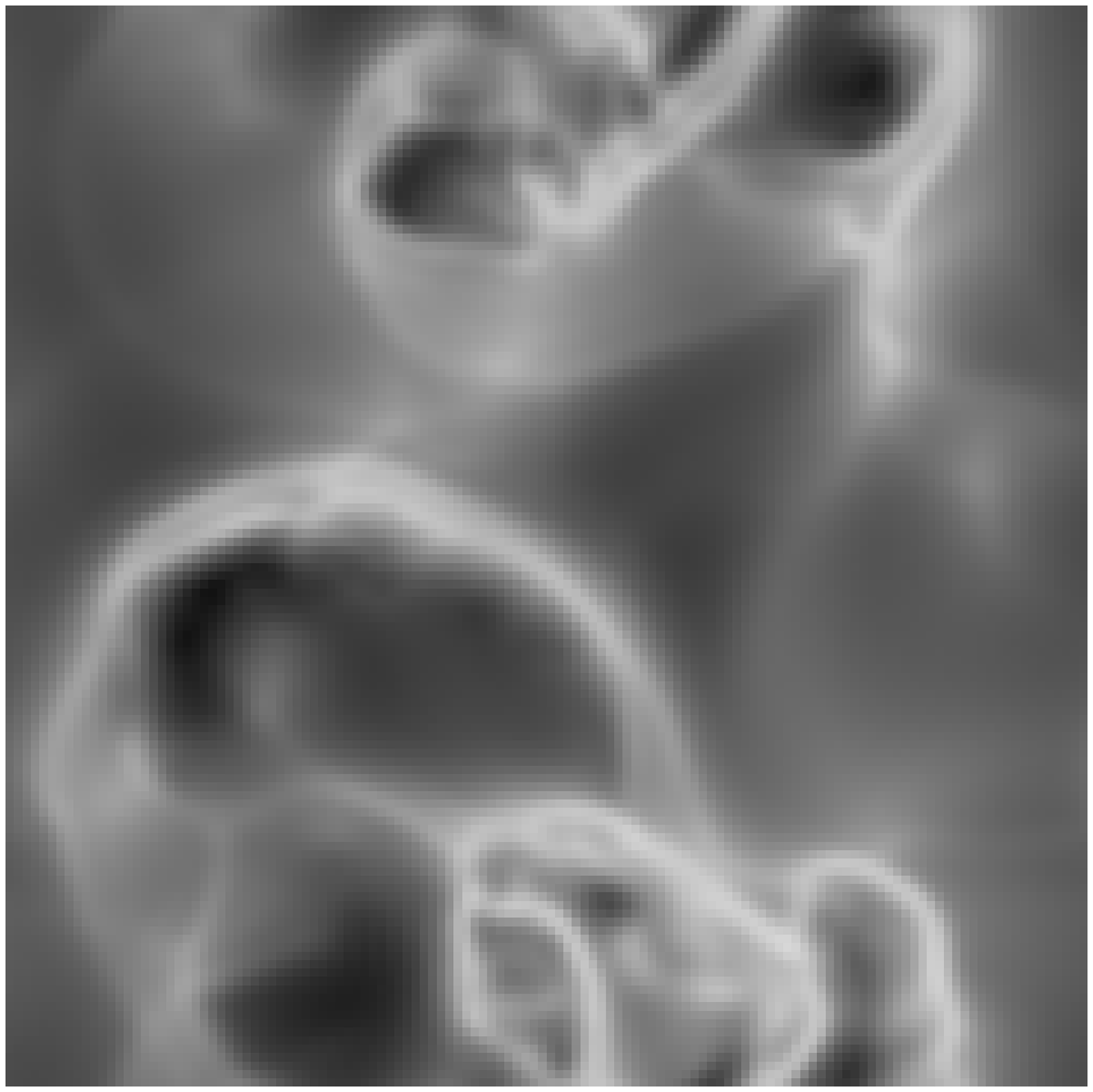}
\caption{Images of the log of the density field of run 6, which used
density-directed star formation (SF), self-gravity and the Coriolis
force, shown at $t=15.0$
Myr ({\it left}) and at $t=149.6$ Myr ({\it right}). SF started at
$t=5.2$ Myr in this run. In the left panel 
small expanding shells are seen, still in their initial
phases. The trapezoidal void near the center was not formed by stellar
activity, but by the turbulent initial conditions. At the later time,
large shells of up to $\sim500$ pc are seen. These are due to induced
SF in the shells. This run exhibits a similar dissipation time as the
fiducial run (30) in spite of the more realistic conditions.}
\label{fig:ISM}
\end{figure}

To this effect, we have also performed a simulation with
these ingredients (run 6, see \S\ref{sec:the_sims}). In fig.\
\ref{fig:ISM} we present images of density field of this simulation
at times $t=14.3$ Myr (left) and $t=144.3$ Myr (right).
At the earlier time, small shells are seen, whose sizes are 
comparable to the sphere of influence of the ``winds'' 
themselves (of radius $\lf=30$ pc). 
At the later time, the shells are seen to be larger, but their 
larger sizes are not only a consequence of the inertial motions 
induced by the forcing, but also of induced-SF events in 
the shells themselves. Even though for this run we cannot estimate
$\uf$ precisely, and consequently $\epsdot$, we can measure the total
injection rate and, from simple visual inspection, we see
that the average radius of the shells 
in the simulation is at least twice as large as $2\lf$. Moreover, on
average we measure $\Ek\approx 0.18$ (in code unities),
$\urms\approx 7$ \kms, and $\td\approx 18$ Myr. We thus see that
this run is in a very similar regime as runs 30 and 35, and in 
particular the dissipation time is within 15\% of that measured 
in these runs, and is 30\% larger than the lower-bound estimate 
for the solar neighborhood given in \S \ref{sec:theory}. Thus, this 
run reassures us that the results obtained with the random placement 
of the stellar sources are applicable in more realistic situations.

\section{Dissipation in decaying turbulence} \label{sec:decay}
 
\begin{figure}
\plottwo{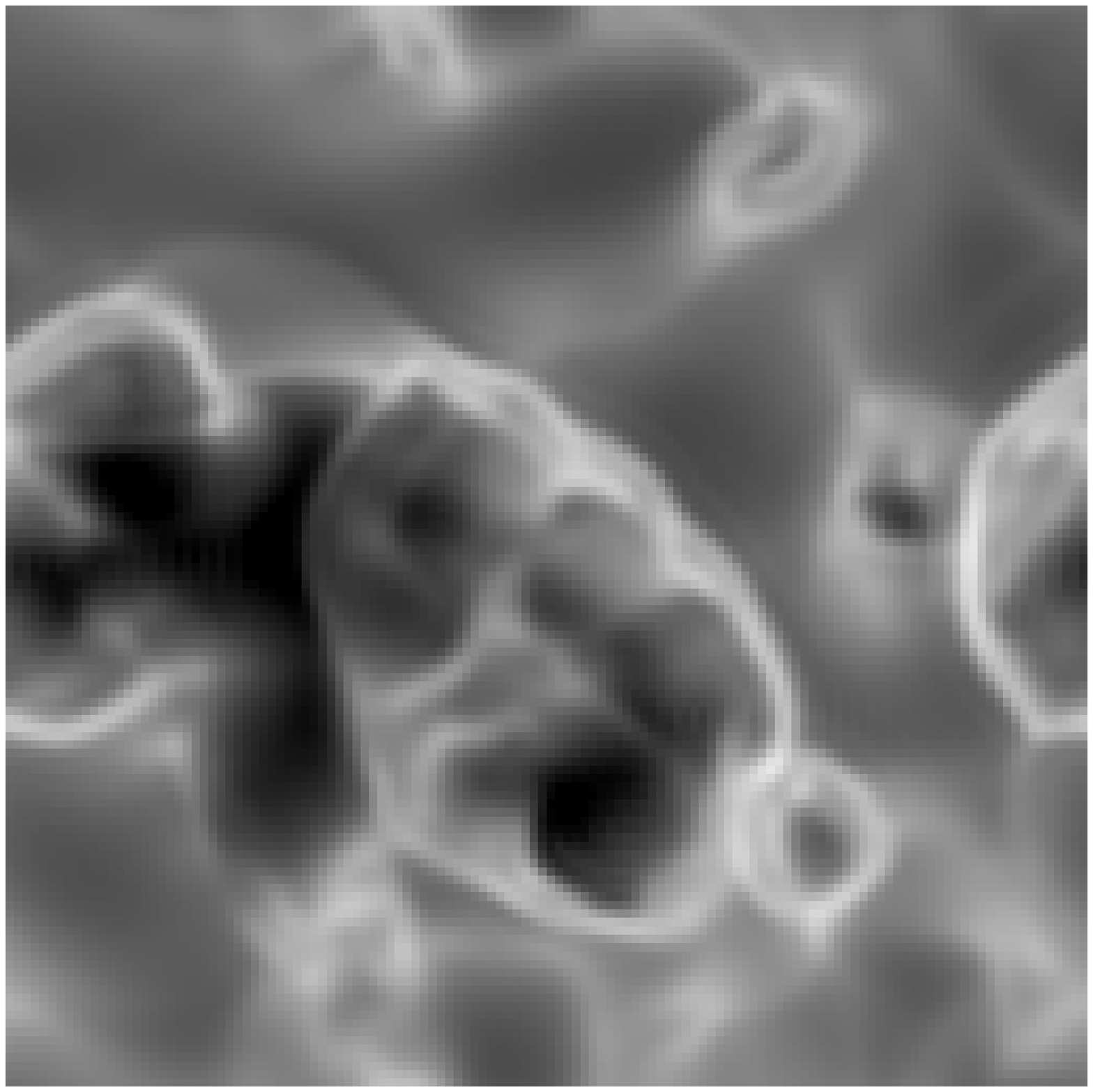}{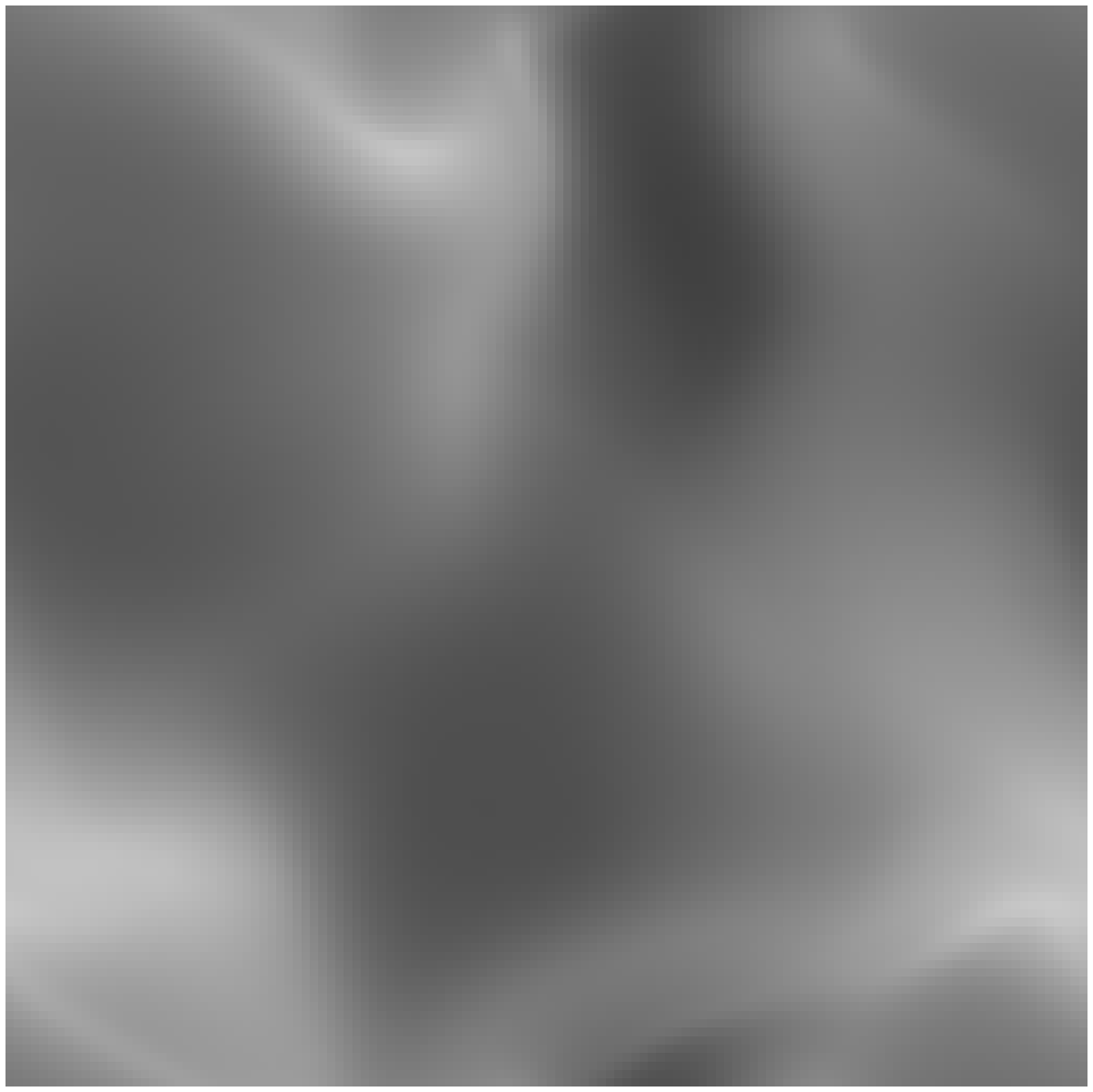}
\caption{Log of the density field for the decay simulation, run 12, at
times $t=0$ ({\it left}) and $t=189.0$ Myr ({\it right}). This
runs is a restart of run similar to run 6 at $t=49.0$ Myr, with
SF, the Coriolis force and self-gravity turned off. 
Expanding shells can still be seen at the earlier time, while a much
smoother density structure is seen at the later time.}
\label{fig:imag_dec}
\end{figure}

\begin{figure}
\plotone{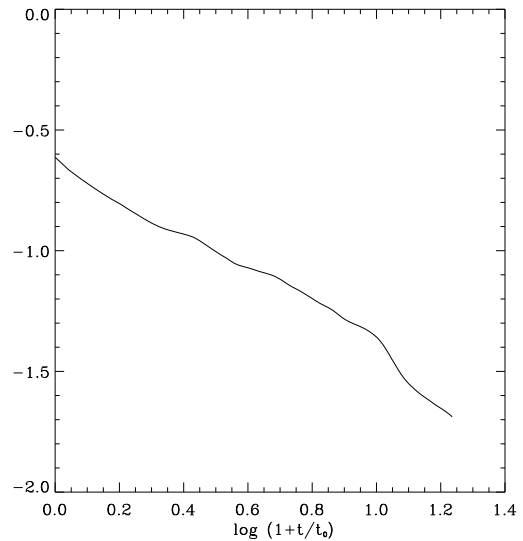}
\caption{Evolution of the total kinetic energy (in code units) for the 
decaying run (run 12).}
\label{fig:evoldec}
\end{figure}

In order to study the decay of turbulence far from the injection
sites, we now consider a simulation in a decaying (i.e., without energy 
injection) regime. To this end, we consider a continuation of run 6, named
run 12, but with the SF turned off. In fig.\ \ref{fig:imag_dec} we
present images of the density field from this run at two
times, one immediately after turning off the SF ($t=0$) and the other
at a much more advanced stage ($t=180.7$ Myr). The former time shows
a large number of expanding shells, while the latter contains no shells
anymore. The evolution of the kinetic energy for this
simulation is shown in fig.\ \ref{fig:evoldec}. We find that $\Ek$
decays roughly as 
\begin{equation} 
\Ek(t)=E_{0}\Bigl(1+\frac{t}{t_0}\Bigr)^{-\alpha}
\label{eq:e(t)}
\end{equation}
with $\alpha \approx 0.8$, in reasonably good agreement with previous
studies of decaying MHD turbulence for isothermal flows (Mac Low et
al. 1998, Stone et al. 1998), and $t_0$ is the code time unit. A
characteristic decay time can be defined in this case as the time
$\Ek$ to decay by a factor of two at the beginning of the simulation. 
For $\alpha \approx 0.8$, this occurs at $t=1.38\times t_0=18$ Myr.

We can explore the decay of turbulent motions
somewhat further. This analysis is greatly simplified in the case of
long times ($\gg t_0$), which allows us to drop the term 1 inside the
parenthesis in eq.\ (\ref{eq:e(t)}), writing
\begin{equation}
\Ek(t)\approx E_{0}\Bigl(\frac{t}{t_0}\Bigr)^{-\alpha}
\label{eq:e(t-1)}
\end{equation}
This limit is reasonable, since
we are interested in the behavior of the turbulence far from the
injection sites. Then, assuming that the rms velocity
$\urms$ scales roughly as the square root of the total kinetic energy
(see Fig. \ref{fig:EvsU}), from eq.\ (\ref{eq:e(t)}) we can write:
\begin{equation} 
\urms(t)\approx u_1\Bigl(\frac{t}{t_0}\Bigr)^{-\alpha/2},
\label{eq:u(t)}
\end{equation}
where $u_1$ is the rms velocity dispersion at $t=t_0$ ($\approx 7$ \kms). 
Similarly to the forced case, 
here we may formally define the dissipation time scale as 
the ratio between $\Ek$ and $|\dot{E}_k|$. Thus, we obtain:
\begin{equation} 
\td(t)\equiv \frac{\Ek}{|\dot{E}_k|}\approx \frac{t_0}{\alpha}
\Bigl(\frac{t}{t_0}\Bigr).
\label{eq:td(t)}
\end{equation}
Thus, {\it the characteristic dissipation time increases
linearly with time in the decaying case}, implying that the
instantaneous dissipation time of the flow depends on the net energy
content of the flow, {\it being smaller for larger energy
contents}. 

Identifying now the decaying case with the regions far from the
injection sites in a forced case with small-scale, spatially
intermittent forcing, the 
temporal decay of the energy can be translated into a decay with
distance. We can estimate the distance ``reached'' by the turbulent
motions in a time $t$ as 
\begin{equation} 
\ell (t)\approx \urms(t)t\approx \ell_0
\Bigl(\frac{t}{t_0}\Bigr)^{1-\alpha/2},
\label{eq:l(t)}
\end{equation}  
where $\ell _0=u_1 t_0 \sim10^2$ pc is the typical length traveled by
a fluid parcel at the characteristic time and velocity at the
beginning of the decaying regime. This length does not exceed the size
scale of the expanding shells produced by the forcing.
Solving for $t$ in eq.\ (\ref{eq:l(t)}) and substituting in eqs.\
(\ref{eq:e(t-1)}) and (\ref{eq:u(t)}), we obtain the dependence 
of $\Ek$ and $\urms$ with $\ell$:
\begin{equation} 
\Ek(\ell) \approx E_{0}\Bigl(\frac{\ell}{\ell_0}\Bigr)
^{-\frac{2\alpha}{2-\alpha}} \label{propagation_eq}
\label{eq:e(l)}
\end{equation}
\begin{equation}
\urms(\ell) \approx u_1\Bigl(\frac{\ell}{\ell_0}\Bigr)
^{-\frac{\alpha}{2-\alpha}}
\label{eq:u(l)}
\end{equation}
For $\alpha \approx 1$ this implies that $\Ek$
and $\urms$ decay with distance as $(\ell/\ell_0)^{-2}$ 
and $(\ell/\ell_0)^{-1}$, respectively.
Thus, ``residual'' turbulent motions with an rms velocity of
roughly 6-10 \kms, as observed in our and other disk galaxies, would
decay to roughly 1/10 of that value on distances $\sim0.6-1$ kpc, in the
absence of energy sources. Note however that, since the decay law
(\ref{eq:e(t-1)}) is a power law (rather than, say, an exponential), the
time required for $\Ek$ to decrease by consecutive constant factors
increases as the the kinetic energy content of the flow decreases
(cf.\ eq.\ [\ref{eq:td(t)}]). This applies also to the decay
with distance, and thus the distance required for the rms velocity to
decrease by a factor of 10 depends on the initial energy content of the
medium. This is also why energy is dissipated so efficiently near the
injection sites.

\section{Effects of resolution and dimensionality}
\label{sec:resol_dim}

For economy, and to maximize our ability to explore parameter space,
in this paper we have limited ourselves to low-resolution, 2D
numerical simulations. However, even though in this paper we are not
attempting to give detailed quantitative results, but only first-order
estimates, it is important to test that even these can indeed be 
attained with the simulations we have employed. 

The two-dimensionality poses an apparently serious problem, since it
is well known that, {\it in the incompressible case}, the energy
dissipation rate has a fundamentally different behavior in 2D
with respect to 3D. While in 3D this rate is
believed to remain finite even in the limit of vanishing viscosity, in
2D it does tend to zero in this limit (see, e.g., Lesieur 1990, sec.\
IX.3.5). Moreover, in 2D the kinetic energy cascades in an ``inverse''
way, from small to large scales (always in the incompressible
case). All of this questions the validity of using 2D simulations for
estimates of the dissipation rate and, in particular, implies that
convergence tests may be meaningless in 2D since the kinetic energy
dissipation rate does not converge as the resolution is
increased and the dissipation coefficients are decreased.

Fortunately, this problem may not be present in the highly
compressible case, in which shocks dominate the dissipation rate, since
it is well known that shocks cause direct transfer from all scales to
the dissipation scales (Kadomtsev \& Petviashvili 1973). This
process is independent of the dimensionality of the flow since the
shocks are essentially one-dimensional structures. Moreover, the
dissipative nature of the shocks suggests that the problem of
vanishing dissipation in the limit of vanishing viscosity in 2D is
eliminated in their presence. This suggestion is supported by
the fact that in our forced simulations the dissipation rate follows
the injection rate closely. In summary, we are confident 
that in the highly compressible 2D case, the dissipation rate can still
be meaningfully measured, and that convergence tests can be performed.

To this effect, we have produced a higher-resolution simulation in
the decaying regime, at $512^2$ grid points, still in 2D, labeled
run 17. In this run, we 
have reduced the hyperviscosity coefficient by a factor of $4^7=16384$, and
the second-order viscosities and diffusivities by a factor of $4^1=4$,
in an attempt to cause the dissipative and diffusive scales to be a
factor of 4 smaller than in the $128^2$ runs, i.e., roughly the same
size in grid points. The factors chosen take into account the order of
the hyperviscous and second-order viscous operators (8 and 2,
respectively), and the fact that the mean Fourier amplitude of the
velocity scales as $k^{-1}$ if the kinetic
energy spectrum $E(k)$ scales as $k^{-2}$, as expected in a flow dominated by
shocks (see, e.g., the review by \VS\ et al.\ 2000 and references therein).
As is the case with the decaying run at $128^2$ resolution (run 12),
run 17 is again a restart of a forced
simulation which used the old scheme of placing the stars at the
highest density sites.

\begin{figure}
\plotone{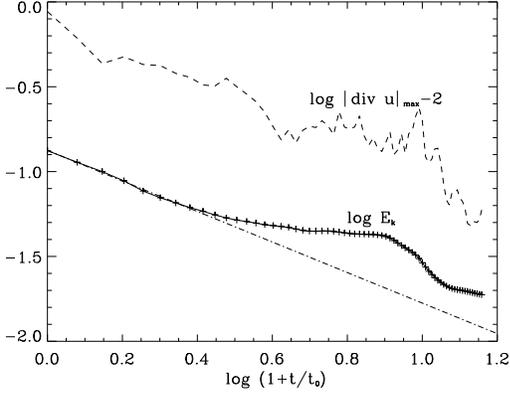}
\caption{Evolution, for run 17 (similar to run 12 but at
resolution 512$^2$), of the log of the kinetic energy
({\it solid line with plus signs}, in code
units) and of the absolute value of the most negative divergence of the
velocity field ({\it dashed line}, displaced by $-2$ in the vertical
direction to fit it in the plot). The dash-dotted line shows a slope
of $-0.9$.}
\label{fig:ek_div_r17}
\end{figure}

Figure \ref{fig:ek_div_r17} shows the evolution of the kinetic energy $\Ek$
for run 17. It is 
seen that over the early epochs of the simulation ($\xi \equiv \log 
(1+t/t_0) \lesssim 0.4$), the energy varies
nearly as a power law with time, with exponent $\sim -0.9$, in
rough agreement with the $-0.8$ exponent observed for run 12, and
also with the exponents reported by Mac Low et al.\ (1998) and Stone
et al.\ (1999). However, at later times, a departure from this
behavior is observed, with the decay rate of $\Ek$ monotonically
decreasing for $0.4 \lesssim \xi \lesssim 0.9$, indicating that at
the higher resolution  of run 17
the decay rate indeed becomes much smaller after the shocks have
subsided. To support this interpretation, in the same figure we also show
the evolution of the most negative value of the divergence of the
velocity field (shown in log of its absolute value). It is seen that
the departure from a power-law decay rate coincides with a drop of the
divergence by nearly a factor of 3 at $\xi \sim 0.5$. Moreover, a
sharp drop in $\Ek$ at $\xi \sim 0.95$ coincides with a large peak in 
the divergence, after which the latter decreases even further and
$\Ek$ returns to a slow decay rate ($\xi \sim 1.1$). In summary,
Fig.\ \ref{fig:ek_div_r17} supports the view that in 2D the dissipation
rate follows 
roughly the same decay rate as 3D simulations (Mac Low et al.\ 1998;
Stone et al.\ 1999) as long as the shocks dominate the dissipation, giving us 
confidence that our 2D simulations provide realistic measurements of
the dissipation rate in the forced regime, which contains large
numbers of shocks, and in the early epochs of the decaying regimes.

Concerning the forced case, a potential problem is that for some 
of the low-resolution runs, the injection scale overlaps with the
dissipation scale (J.\ Brasseur 2000, private communication), and thus
there is the risk that this energy never reaches the larger scales
of the flow. We believe this is not really a problem, because, as
mentioned above, shocks ``short-circuit'' the energy directly from
the energy-containing scales to the dissipations scales, so the
cascade through intermediate scales is 
eliminated anyway. In physical space, what happens is that, even if the
forcing has characteristic scales larger than the dissipation scale,
its effect is to induce shocks into the medium, which are small-scale 
structures. Indeed, runs 40 and 41, which have a value of $\lf$ 
twice as large as that of run 30, still agree with our model
prediction within $\sim45$\% and $30$\%, respectively. 

Nevertheless, it is important to verify that the agreement between our 
semi-empirical model of \S \ref{sec:theory} and the low resolution 
simulations is maintained in higher-resolution runs. To this effect,
we have performed a $512^2$ forced run, labeled run 44 (Table 1) in
which the physical size of the box (1 kpc) is maintained, but the smallest
resolved scale is now 1.95 pc. This run
uses the same viscosities and difussivities as run 17, and its parameters are
chosen so that it is a rescaling of run 30 (the fiducial run): $\lf$
is 4 times larger in pixels, so it keeps the same physical size in
parsecs. However, the dissipation scale is
four times smaller in physical size, becoming safely smaller
than the injection scale. All other parameters are the same as in run
30. From Table 2, it can be seen that run 44 conforms nicely to our
semi-empirical model, its predicted and measured injection times 
agreeing within $\sim15$\%.

\section{Discussion} \label{sec:discussion}

\subsection{Comparison with previous work} \label{sec:comparison}

The main difference between our MHD simulations of
supersonic compressible flows and previous
ones aimed at exploring turbulence dissipation resides in
the method we use to drive the turbulence. As pointed out in \S
\ref{sec:SF},
we have tried to represent the way in which kinetic energy is injected
into the ISM of disk galaxies by stellar
sources. Moreover, since
our interest in this paper focuses in the large-scale ISM, we have
taken parameters of the flow corresponding to
the solar neighborhood, as done in previous papers modeling the ISM
at large (Chiang \& Bregman 1988; Rosen \& Bregman 1995; \VS\ et al.\
1995, 1996; Passot et al.\ 1995), and have {\it not} assumed
the flow to be isothermal. All of this contrasts with recent studies of
turbulence dissipation in molecular clouds (Mac Low et al. 1998;
Stone et al. 1998; Padoan
\& Nordlund 1999; Mac Low 1999a), which have considered isothermal
flows with large-scale, ubiquitous forcing. On the other hand, those works
have been based on 3D simulations, while
here we have used 2D simulations. However, large differences
on the dissipation rates and timescales between 2D and 3D
simulations are not expected (see e.g., Stone et al. 1998 and \S
\ref{sec:resol_dim} above).
Specifically, the kinetic energy dissipation rates             
in 3D are expected to be slightly higher than in 2D,
particularly in the regime of decaying turbulence,
because at higher dimensionality, more shocks are produced
due to the larger number of degrees of freedom (e.g.,
Mac Low et al. 1998). For example, Stone et al. (1998) reported that
the energy decay times found for $2+1/2$D MHD
simulations are a factor of 1.50---1.75 times larger than
those for 3D.

\begin{figure}
\plottwo{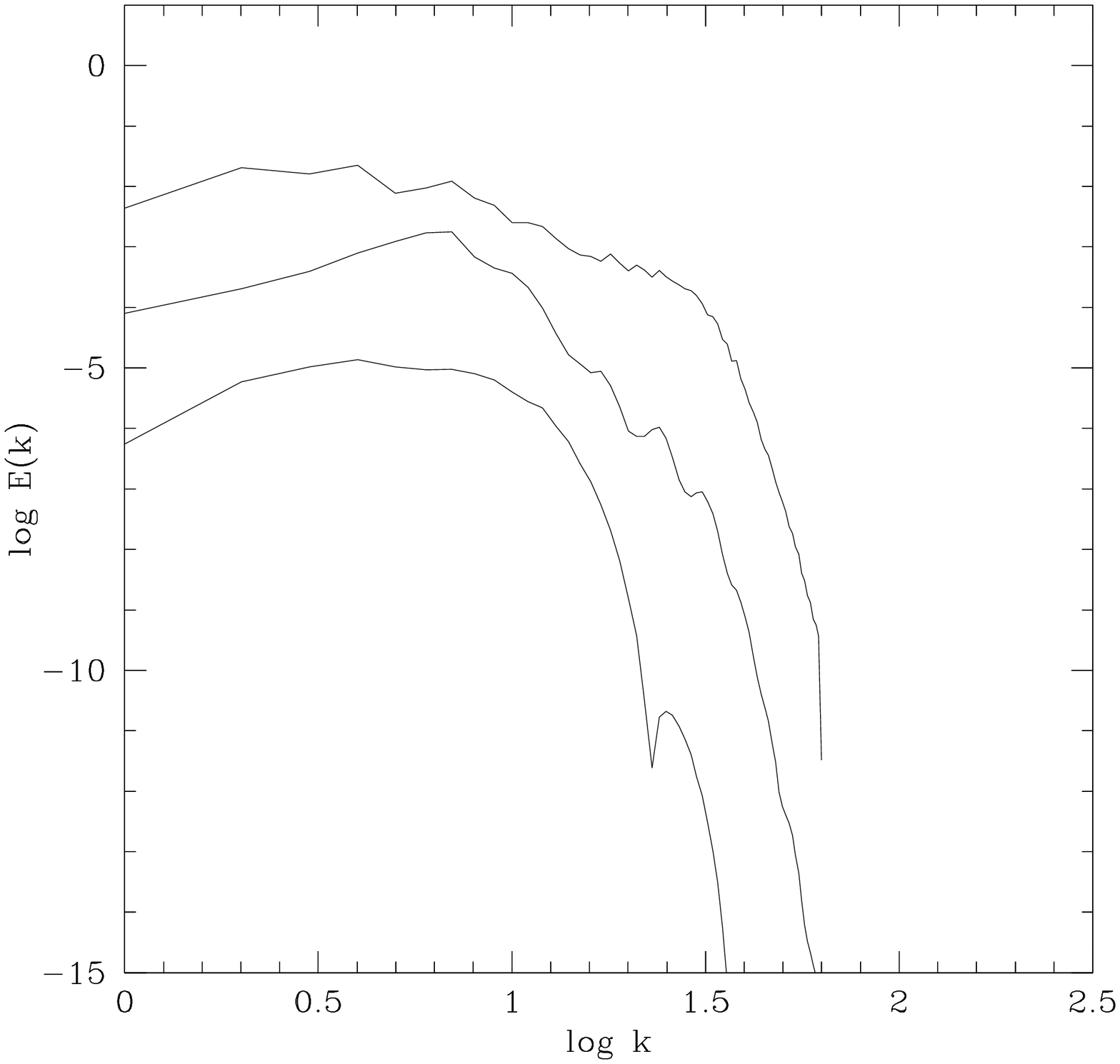}{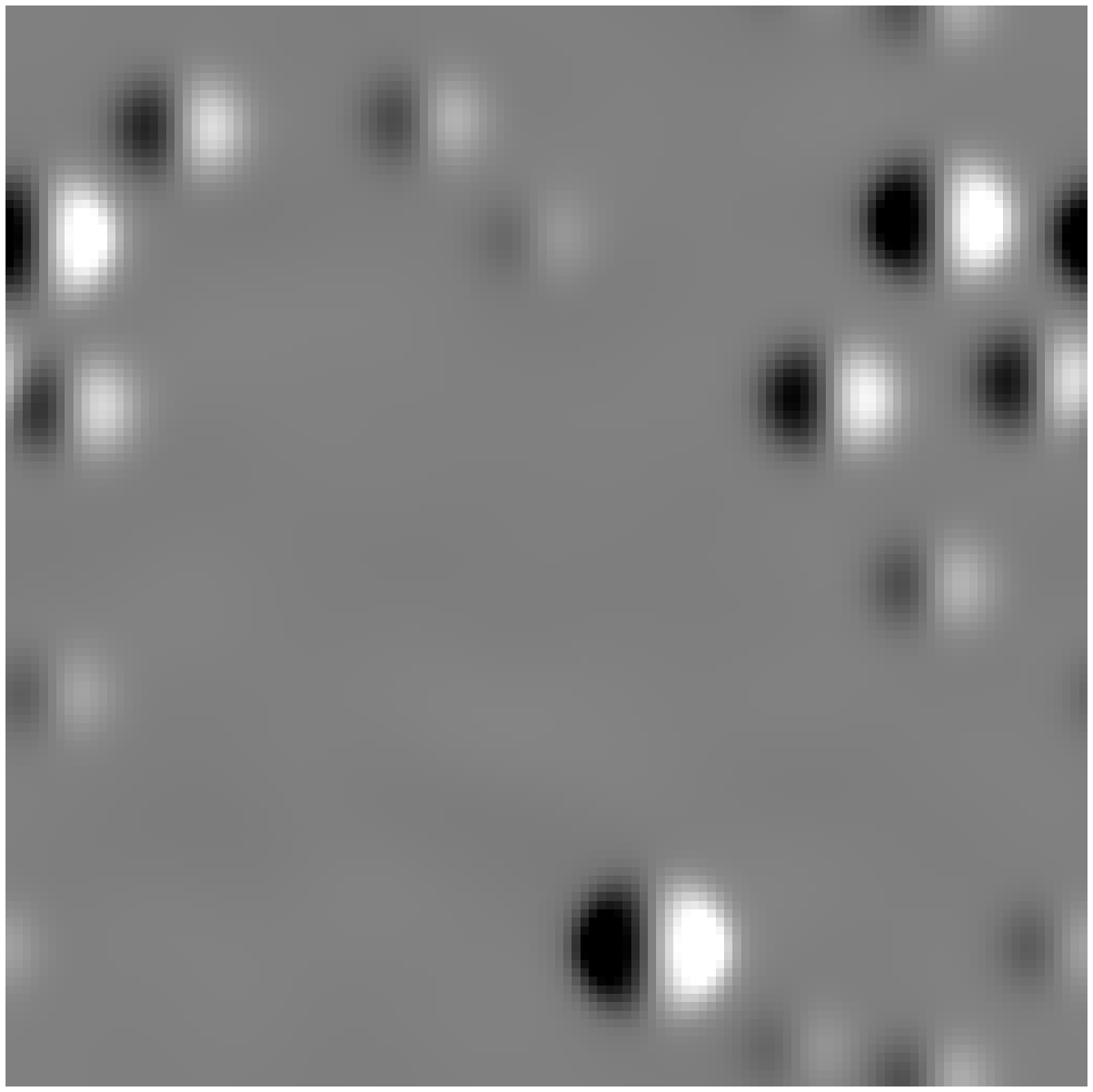}
\caption{{\it a)(Left)} Velocity spectra of the fiducial run at times
$t=1.36$ Myr {\it 
lower line}, $t=6.8$ Myr ({\it middle line}) and $t=136$ Myr ({\it
upper line}). {\it b) (Right)} $x$-component of the velocity field at
$t=6.8$ Myr. 
At $t=1.36$ Myr, the spectrum is dominated by the
(mild) initial conditions, but a peak at $\log k = 1.4$ corresponds to 
the first (and only) shell present at that time, just formed. At
$t=6.8$ Myr this peak has moved to $\log k \sim 1.2$ as the shell has
expanded, and several other ripples at larger $k$ correspond to new
shells. At lower $k$ the spectrum corresponds to the {\it ensemble} of 
shells. At $t=136$ Myr the spectrum has reached a power-law shape,
indicating fully developed turbulence.}
\label{fig:spec}
\end{figure}

At the results level, in spite of the very different physical
regimes we consider, we agree with previous works in
that {\it compressible MHD flows dissipate their
turbulent kinetic energy very efficiently}. More specifically, for
their molecular cloud simulations, Stone et al. (1998)
and Mac Low (1999a) find that the dissipation time is of the
order to the flow crossing time at the energy-containing scale
$\lambda_{\rm E}$, i.e. $\td\approx \lambda_{\rm E}/\urms$. However,
in their work, this scale is very close to the injection scale because of the
large-scale and ubiquitous nature of the injection. In our case, on the
other hand, $\lambda_{\rm E}$ is in general quite larger than the driving
scale $2\lf$, because of the expansion of the shells,
which is a way of transferring
the energy to larger scales, and possibly also because energy that is not
dissipated in shocks may escape towards the largest scales through an inverse
cascade due to the two-dimensionality (to be investigated elsewhere). Indeed,
fig.\ \ref{fig:spec}a shows the energy spectrum for run 30 at times
$t=1.36$, 6.8 and 136 Myr. For reference, fig.\ \ref{fig:spec}b shows
the $x$-component of the velocity field at $t=6.8$ Myr.
It can be seen that at $t=1.36$ Myr, the spectrum contains only one
peak at $\log k\sim 1.4$ (the bulk of the spectrum at lower $k$ at this time
corresponds to the very mild initial conditions with $\urms \sim 0.1$ km 
s$^{-1}$). A scale (in pixels) corresponding to wavenumber $k$ can
be defined as $\lambda = 128/k$. For $\log k=1.4$, $\lambda \approx 5$ 
pixels, and indeed at this time there is only one expanding shell in the
simulation with ``radius'' $\sim 4$ pixels, measured as the distance
from the center to the velocity maximum of the shell (this distance
has been measured from cuts through the $x$-component of the velocity
passing through the shell center, similar to those in fig.\
\ref{fig:au_profiles}, not shown). 

At $t=6.8$ Myr, this peak has moved to $\log k \sim 1.22$ ($\lambda 
\approx 7.7$ pixels) due to the expansion of the shell, whose radius has 
grown to $\sim 6$ pixels. New peaks at larger $k$ have
appeared corresponding to the formation of new sources. The maximum of the
spectrum (roughly, the ``energy-containing scale'') at $t=6.8$ Myr, 
at  $\log k \sim 0.8$, corresponds to the diameter of the largest 
shell at that time. Energy at even larger scales must correspond 
to the size of the region spanned
by the various sources at that time, since the shells themselves have
not grown further than $\log k = 0.8$ yet. By  $t=136$ Myr, the spectrum
is very close to a power law in the range $0.6 \lesssim \log k
\lesssim 1.5$, suggesting the existence of fully developed turbulence
at that time.

Since the results above suggest a clear separation between the
injection and energy-containing scales in our simulations, in order to
compare our results to the works of 
Stone et al.\ and of Mac Low, we can quantitatively estimate the
energy-containing scale in our simulations as the scale corresponding to
the centroid of the energy distribution, given by (see, e.g., Novikov 1978;
Santangelo et al.\ 1989; McWilliams 1990)
\begin{equation}
\lambda_{\rm E} \equiv 2 \pi \Bigl(\frac{\sum_\k k |\u_\k|^2}
{\sum_\k |\u_\k|^2}\Bigr)^{-1},
\label{eq:centroid}
\end{equation}
where $\u_\k$ is the Fourier amplitude of the mode with (vector)
wavenumber \k.
Table 2 gives the values of $\lambda_{\rm E}$ for all runs at $t=136$
Myr, a time 
at which the spectrum has become essentially stationary, and the value of
the large-scale crossing time 
$t_{\rm cr}\equiv \lambda_{\rm E}/\urms$, calculated
with the rms velocity at that same time. It is seen that
the crossing times do not correlate well with the measured
dissipation times, and are in general quite larger. We
speculate that this is due to the fact that $t_{\rm cr}$ measures
essentially the dissipation time at the largest scales, and fails to
capture the local dissipation near the sources. In particular, the
local and large-scale dissipations contribute in different proportions
to the total rate depending on the size, number and specific energy
injection rate of the sources, and thus the crossing time is not a             
good estimator of the true injection and dissipation times in the case
of non-ubiquitous, small-scale forcing.

In conclusion, we do not expect that a simple law of the type
$\td=\lambda_{\rm E}/U$, with $U$ either the rms of the characteristic 
forcing velocity, can hold in the case with non-ubiquitous forcing:
another parameter, the filling factor of the forcing regions, is
crucial, as it determines the total amount of energy injected by
sources of given local parameters, the proportion in which $\uf$ 
contributes to $\urms$ and the amount of overlapping between the regions.

\subsection{Implications for models of galaxy evolution} \label{sec:gal_evol}

From the point of view of galactic evolution, the global ISM
processes, which take place at large space and time scales, are the
most relevant. As mentioned in the Introduction, several galactic
evolution models are based on the idea that the SF rate is controlled
by a disk vertical balance in the ISM between the rates of energy injection due to 
SF and energy dissipation. Since turbulent kinetic energy is believed 
to be one of the dominant sources of pressure in the ISM, the relevant
dissipation rate for these models is that of turbulence, motivating
our study.  We have shown that $\Ek$ is dissipated very efficiently 
near the input sources (locally). This implies that very large ``active'' 
turbulent zones can exist in galaxies only if the forcing regions are 
themselves large, as a consequence of either very powerful sources, or
strong clustering and self-propagating SF. This might be the rule for
starburst galaxies, but generally the exception in normal disk
galaxies. Besides, as
results from our simulations have shown (\S\ref{sec:driven_turb}), the 
dissipation time $\td$ indeed increases as the source size and filling 
factor increase.

In the energy balance in the vertical direction assumed in the 
galaxy evolution models mentioned above (taking a given
energy input rate, given by the SF rate, and estimating the
dissipation rate as $\propto \urms^2/\td$), the equilibrium gas velocity
dispersion is determined by $\td$. Hence,
assuming vertical (one-zone) hydrostatic equilibrium, the disk HI 
thickness will depend basically on $\td$. According to the results
of these galaxy evolution models (e.g., Firmani et al. 1996; Avila-
Reese \& Firmani 2000; Firmani \& Avila-Reese 2000), a realistic
HI gas thickness at the solar neighborhood is obtained with values of
$\td$ roughly twice as large as the values found here (all the
other relevant quantities as gas surface density, circular velocity and
star formation rate, are self-consistently calculated and
agree well with observations). From the hydrostatic equilibrium condition,
indeed it is easy to see that a turbulent gas layer supported by a velocity
dispersion $\approx9$ \kms\ would have a thickness roughly 3
times smaller than observed, suggesting that probably other kinds
of pressure, such as the magnetic field and cosmic-rays, are also important 
for supporting the galactic gas layer.

On the other hand, it is important to have in mind that the
vertical galactic disk is actually stratified; as the gas density
decreases, the survival and propagation of turbulent motions could
be easier, and therefore $\td$ would be longer. In fact, shells are
well known to accelerate when they transit from a high- to a
low-density medium. In this sense, the dissipation
times calculated here for a fixed gas average density
can be interpreted as a lower limit; hence, the disk thicknesses 
estimated with our turbulent dissipation times will be also a lower 
limit. In any case, according to the results of galaxy evolution models,
with self-regulated SF, the dissipation time obtained here is not too far
from the values necessary to reproduce several observational data.
Thus, our results do not disagree with the idea that turbulent 
motions induced by SF feedback contribute significantly to support 
the vertical HI disk (e.g., Lockman \& Gehman 1991; Firmani \& Tutukov 1992, 
1994; Ferrara 1993; Dopita \& Ryder 1994), although some extra support
may be necessary to keep the HI and gas ionized layers up at the
observed heights.

Finally, it is important to remark that the turbulent dissipation
time may change along the galactic disk since the gas densities 
and dynamical conditions change with radius. According to Firmani et
al.\ (1996), $\td\propto R$ when the rotation curve is flat. 
In \S\ref{sec:theory} we obtained a theoretical expression for 
$\td$ (actually $\tin$; eq.\ [\ref{eq:obs1}]) which in particular implies a 
dependence on galactic radius through $\langle \Sigmag \rangle$
and $\sfrOB$.
The OB-star formation rate is directly proportional to the 
total formation rate $\sfrS$ and, in most disk galaxies,
$\sfrS\propto \Sigmag^n$, with $n\approx 1.4-2.0$ (e.g.,
Kennicutt 1998 and references therein). Therefore, from
eq.\ (\ref{eq:obs1}) one obtains that $\tin (r)\propto
\langle \Sigmag \rangle ^{n-1}(r)$. The gas surface density
typically decreases with radius. For example, in the Galaxy,
from 4 to 16 kpc, $\Sigmag$ decreases roughly as $r^{-1}$
(Dame 1993). Hence, the dissipation time should decay as the
Galactocentric radius $R$ to some power, probably less than 
unity.
 
\subsection{Implications for models of galaxy formation} \label{sec:gal_form}

Several approaches of galaxy formation within the context of 
the hierarchical CDM-based scenario ---in particular the so-called
semi-analytical models (e.g., Kauffmann et al.\ 
1993; Cole et al.\ 1994; Somerville \& Primack 1999; van den Bosch 
1999)--- require the feedback from SF to be able to reheat the 
cooled gas up to the virial temperature of the whole
dark matter halo. This requirement implies that a significant fraction  
of the released SN energy (probably more than 10\%) remains as kinetic 
energy and that this energy is not dissipated within the disk. 

The physical picture invoked in the semi-analytical models 
implies a self-regulated SF not at the level of the 
{\it disk} ISM, but at the level of the hypothetical {\it intra-halo} medium. 
The SNe formed in the disk are assumed to reheat the cold disk 
gas up to the virial temperature of the cosmological dark
matter halo, {\it driving it back} into 
the intra-halo medium, and occasionally expelling it completely 
from the system (White \& Frenk 1991). Thus, a crucial question
for models of galaxy formation is whether the energy 
released by SNe and stars in the disk is able not only to maintain
the warm and hot phases and the stirring of the disk ISM, but 
also to maintain a huge hot gas corona in quasi-hydrostatic
equilibrium with the cosmological dark halo (the intra-halo medium); 
the estimated sizes of these dark halos are several tens
and hundreds of kiloparsecs for dwarf and giant galaxies,
respectively. Note that this intra-halo medium should not be confused
with the diffuse ionized gas and high-velocity-dispersion HI gas at scale 
heights of one or a few kiloparsecs above the disk plane, also
frequently referred to as a ``halo'' in the ISM community
(see e.g., Reynolds 1997; Kalberla \& Kerp 1998; 
Mac Low 1999b). 

The question is then whether the intra-halo gas can be dynamically heated 
by the turbulent energy input due to stellar winds produced by
SNe and OB stars. Indeed,
numerical simulations show that up to $5-10$\% of the energy
produced by SNe can be transformed into kinetic energy in the
extreme case of multiple SN explosions forming supershells 
(e.g., Silich et al.\ 1996; Korpi et al.\ 1999); for isolated
SNe, this fraction is lower than $\sim 1$\%.

However, according to the results obtained here, most of
kinetic energy in these ``active'' 
turbulent regions is dissipated {\it locally} and the $\urms$ of
the ``residual'' turbulent motions quickly decays with distance
(roughly $\urms \propto \ell^{-1}$). In order for the 
turbulent gas to be driven back into the intra-halo medium,
the typical sizes of most of the forcing shells 
should exceed the gaseous disk height (supershells). This
could be a common situation in dwarf starburst galaxies, but it 
is not the rule in normal disk galaxies. In fact, even in dwarf
starburst galaxies, gas ejection from the disk seems to 
be not very efficient, according to recent hydrodynamical
simulations by Mac Low \& Ferrara (1999). Besides, numerical 
simulations have shown that the disk magnetic field is 
an efficient shield able to confine most superbubbles 
preventing immediate blowout (e.g., Slavin \& Cox 1992;
Franco et al.\ 1995; Tomisaka 1992, 1998). 
Therefore, we conclude that {\it it is
very unlikely that the turbulent kinetic energy injected by SNe
and massive OB stars into the disk ISM can survive, propagate 
and drive back the gas into the intra-halo medium} as the
semi-analytic models of galaxy formation require.

We should note that our results are for densities typical 
of the ISM in the disk plane at the solar neighborhood. One may 
think that the disk stratification might reduce the dissipation 
rate so much that the turbulent kinetic energy could actually 
reach the intra-halo medium. However, since the HI disk scale 
height is larger than the typical dissipation lengths we have found, 
we believe any escaping energy will be too weak to provide support
to the intra-halo medium. 
Simulations in a stratified medium will be 
presented elsewhere, with a code that does not require periodic
boundary conditions.
 
An important question not treated in this paper is where 
the turbulent dissipated energy goes. It is well known that
shocks efficiently transform kinetic energy into thermal energy. Thus, 
the heated gas at relatively large heights could be a source 
of high-energy photons able to ionize the low density extraplanar
medium or even the hypothetical intra-halo medium.
For example, Slavin (2000) has showed
that hot gas in cooling SN remnants can be an important source
of photoionization for the diffuse ionized gas. Dissipation in 
turbulent mixing layers has been also proposed as a mechanism 
to produce energetic radiation (Slavin, Shull, \& Begelman 1993).
Minter \& Spangler (1997) have found that turbulent dissipation
heating like that due to the ion-neutral collisional damping in
the fluid-like turbulence model is able to provide a substantial
contribution to the energy budget of the diffuse ionized gas.   
More studies on dissipation of turbulence and the thermo-hydrodynamics
of the extraplanar and intra-halo medium are certainly necessary.

\section{Concluding remarks}
\label{sec:concl}

\subsection{Summary} \label{sec:summary}

We have used 2D simulations of MHD, compressible, self-gravitating
flows in the presence of parameterized heating and cooling and
wind-like stellar kinetic energy input in order to study 
turbulence dissipation in the ISM. The ``winds'' are
described by the typical  radius $\lf$ of their region of influence and the
total kinetic energy $\Es$ they inject into the medium. These
parameters and the OB star formation rate $\sfrOB$ completely  
characterize the forcing mechanism. It is important to
emphasize that $\lf$ is {\it not} the size ultimately reached by the
expanding shells, but the scale over which the force is directly
applied. The remainder of the shell expansion is inertial and strictly
speaking does not constitute a forcing. Our main conclusions and
implications are:

\noindent
1. The non-ubiquitous, small-scale nature of the
forcing implies that the forced regions have small filling factors,
giving rise to the coexistence of both forced and decaying turbulent 
regimes within the same flow.

\noindent
2. The kinetic energy injection and dissipation
rates are always very close to each other, strongly suggesting that most of
the dissipation occurs at or near the sources, where
shocks are common, and that, for practical purposes, the injection and
dissipation times are equal. In general, the flow dissipates its turbulent 
energy rapidly, in roughly 15--20 Myr.

\noindent
3. In the forced regime, the {\it global} dissipation timescale $\td$ 
is empirically found to depend mainly on the forcing scale $\lf$, and
only very weakly on $\Es$ (or $\epsdot$, the
specific power injected per source) and $\sfrOB$. 

\noindent
4. Expressing $\tin$ in terms of average quantities in the flow and
parameters of the forcing, we showed that $\td\approx \urms^2/(\epsdot
\ f)$, where 
$f$ is the filling factor of the forcing regions. In terms of 
measurable properties of the ISM, 
$\td \gtrsim \langle \Sigmag \rangle \urms^2/(\Es \ \sfrOB)$,
where $\langle \Sigmag \rangle$ is the average gas surface
density. For the solar neighborhood,
$\td \gtrsim 1.5\times 10^7$ yr. Since $\sfrOB\propto \sfrS\propto
\Sigmag^n$, $\td\propto \Sigmag^{n-1}(R)$,  $\td$ is expected to vary
with Galactocentric radius. 

\noindent
5. In the decaying regime, the turbulent kinetic energy decays as
$\Ek(t)\propto (1+t/t_0)^{-\alpha}$, with $\alpha \approx
0.8$. For this value of $\alpha$, the characteristic time for decay to 
half the initial energy is $\sim18$ Myr, but this time depends on the
total kinetic energy content of the flow, because of the power-law
dependence. We associate the decaying
case to the regime existing in regions distant from the energy
sources, which contains the ``residual'' turbulence left over from the
strong local dissipation operating in the neighborhoods of the sources.

\noindent
6. From dimensional arguments, we suggest that the kinetic energy
$\Ek$ and the rms velocity dispersion $\urms$ of the ``residual''
turbulence (far from the forcing regions) should decay 
with distance $\ell$ as $\ell^{-2 \alpha/(2-\alpha)}$ and
$\ell^{-\alpha/(2-\alpha)}$, respectively. For $\alpha \sim1$, $\Ek
\sim \ell^{-2}$ and $\urms \sim \ell^{-1}$.

\noindent
7. Our results imply that energy is dissipated much more efficiently
at larger kinetic energy contents in the flow, in such a way that the
dissipation time is rather insensitive to the total energy content.

\noindent
8. If our results are applicable to the stratified vertical
direction in the Galactic disk, then turbulent motions produced 
near the disk plane will propagate up to distances not too much
smaller than the observed HI disk semi-thickness. This 
is consistent with models of galaxy
evolution where the HI disk thickness is determined mainly by the
turbulent kinetic energy content of the medium, which in turn
results from the balance between kinetic energy injection by stars and
its dissipation rate. However, some extra support is probably
necessary to keep the HI and ionized layers up at the observed
heights.

\noindent
9. Our result of mostly local dissipation, again if applicable in the
vertical direction, is in conflict with models of galaxy 
formation where the turbulent kinetic energy injected by SNe and
stellar winds is
assumed to reheat and drive back the gas from the disk
into the cosmological dark matter halo (of sizes 15-20 times the
disk size) in such a way that SF is self-regulated at the level of the 
whole intra-halo medium. Only in cases
of non-stationary runaway SF (starbursts), most of the superbubbles
might be able to blowout of the disk expelling large amounts of 
gas and energy into the intra-halo medium.   

\subsection{Importance of small-scale, intermittent forcing} 
\label{concl_intermit}

The results of this paper imply that ISM turbulence is essentially 
different from ideal homogeneous turbulence due to the nature of the 
intervening energy injection (forcing) mechanisms. Although several
authors (e.g., Scalo 1987; Fleck 1983; Norman \& Ferrara 1996; see
also the reviews in \VS\
1999; \VS\ et al.\ 2000) have already discussed the fact that  
in ``traditional'' turbulence energy is injected at large scales
while in the ISM it is injected  over a range of scales (and
preferentially at small ones), the consequences of the
spatially- and temporally-discrete nature of interstellar energy
injection have barely been discussed to our knowledge. 

The effects of intermittent forcing have recently begun to be 
considered in other frameworks as well. Bec, Frisch \& Khanin (2000)
have studied the statistics and solvability properties of Burgers 
turbulence in the presence of ``kick'' forcing
(i.e., the forcing is a series of delta functions in time), pointing 
out that such a flow combines features of both purely decaying and
continuously forced cases. L\'eorat, Passot \& Pouquet (1990) and \VS, 
Gazol \& Scalo (2000) have shown that small-scale forcing is able 
to prevent the development of the gravitational and thermal instabilities,
respectively. Finally, Kornreich \& Scalo (2000) have considered the
exposure of starless interstellar clouds to the intermittent passage
of shock waves originated at distant injection sites, to show that
the typical time between passages should be comparable to the decay
time within the clouds, explaining their apparently turbulent state
even in the absence of local energy injection sources.

In conclusion, the discrete, small-scale nature of interstellar kinetic
energy injection seems to be responsible for a new, rich kind of turbulent
flow, whose dissipation properties seem to pose clear constraints on
models of galaxy formation and evolution.

\acknowledgments

This work has benefitted extensively from comments from J.\ Brasseur,
A.\ Brandenburg, E.\ Ostriker and J.\ Stone while one of us (E.V.-S.)
attended the ``Astrophysical Turbulence Program'' of the Institute of
Theoretical Physics of the University of California at Santa Barbara,
and from a thorough referee's report by M.-M. Mac Low.
This work was supported by CONACYT grant 27752-E to E.V.-S.

\clearpage

\begin{center}
REFERENCES
\end{center}

\begin{description}

\item Alfaro, E.J., Cabrera-Cano, J., \& Delgado, A.J. 1991,
\apj, 378, 106

\item  Avila-Reese, V. 1998, PhD. Thesis, U.N.A.M.

\item  Avila-Reese, V., \& Firmani, C. 2000, RevMexAA, 36, 23

\item  Avillez, M. A. 1999,  \mnras, in press

\item Bania, T. M. \& Lyon, J. G. 1980, ApJ, 239, 173

\item Bec, J., Frisch, U. \& Khanin, K. 1999, J. Fluid Mech., submitted
(chao-dyn/9910001)

\item Biskamp , D. 1994, Nonlinear Magnetohydrodynamics (Cambridge
Univ. Press, Cambridge, England)

\item Chiang, W., \& Bregman, J.N. 1988, \apj, 328, 427

\item Chiang, W., \& Prendergast, K. H. 1985, \apj, 297, 507

\item  Cole, S., Aragon-Salamanca, A., Frenk, C.S., Navarro, J., \& Zepf, S.
1994, \mnras, 271, 781

\item Cox, D. \& Smith,  1974, ApJ, 189, L105

\item Dame, T. 1993, in Back to the Galaxy, eds. S. Holt, F. Verter,
(AIP), p. 267 

\item Dopita, M.A. 1990, in The Interstellar Medium in Galaxies, eds. H.A.
Thronson, Jr. \& J.M. Shull (Dordrecht: Kluwer), p. 437

\item Dopita, M.A., \& Ryder, S.D. 1994, \apj, 430, 163.

\item Elmegreen, B.G. 1991, in The Galactic ISM, eds. W.B. Burton,
B.G. Elmegreen, R. Genzel (Springer-Verlag), p. 157

\item Ferrara, A. 1993, \apj, 407, 157 

\item Field, G.B., Goldsmith, D.W., \& Habing, H.J. 1969, \apj, 155, L149

\item  Firmani, C., \& Avila-Reese, V. 2000, MNRAS, 315, 457

\item  Firmani, C., \& Tutukov, A.V. 1992, \aap, 264, 37

\item  Firmani, C., \& Tutukov, A.V. 1994, \aap, 288, 713

\item  Firmani, C., Hern\'{a}ndez, X., \& Gallagher, J. 1996, \aap, 308,
403

\item Fleck, R. C. Jr. 1983, ApJ, 272, L45

\item Franco, J., Santill\'an, A., \& Martos, M. 1995, in The
formation of the Milky Way, G.Tenorio-Tagle, M.Prieto, \& S\'anchez, 
F. eds. (Cambridge Univ. Press), p. 515

\item Friedli, D., \& Benz, W. 1995, \aap, 301, 649

\item Frisch, U. 1995, Turbulence (Cambridge: Cambridge Univ.
Press, Cambridge, England) 

\item Galtier, S., Politano, H., \& Pouquet, A. 1997, Phys.Rev.Lett,
79, 2807

\item Gerritsen, J.P.E. 1997, PhD. Thesis, Groningen University

\item Gerritsen, J.P.E., \& Icke, V. 1997, \aap, 325, 972

\item Hossain, M., Gray, P., Pontius, D., Matthaeus, W., \& Oughton,
S. 1995, Phys.Fluids 7, 2886

\item Kadomtsev, B.B., \& Petviashvili, V.I. 1973, Sov. Phys. Doklady,
18, 115

\item Kalberla, P.M.W., \& Kerp, J. 1998, \aap, 339, 745

\item  Kauffmann, G., White, S.D.M., \& Guiderdoni, B. 1993, \mnras,
264, 201

\item Kennicutt, R.C. 1998, \apj, 498, 541

\item Kolmogorov, 1941, Dokl. Akad. Nauk 30,301

\item Kornreich, P. \& Scalo, J. 2000, ApJ, 531, 366

\item L\`eorat, J., Passot, T. \& Pouquet, A. 1990, MNRAS, 243, 293 

\item Lesieur, M. 1990, Turbulence in Fluids, 2nd ed.\ (Dordrecht:Kluwer)

\item Lockman, F.J, \& Gehman, C.S. 1991, \apj, 382, 182

\item Mac Low, M.-M. 1999a, \apj, 524, 169

\item Mac Low, M.-M. 1999b, in New Perspectives on the Interstellar Medium, 
eds. A. R. Taylor and T. L. Landecker (ASP: San Francisco), in press

\item Mac Low, M.-M., \& Ferrara, A. 1999, \apj, 513, 142

\item Mac Low, M.-M., Klessen, R. S., \& Burkert, A., \& Smith, M.D. 
1998, \prl, 80, 2754

\item McWilliams, J. C. 1990, Phys. Fluids A, 2, 547  

\item McKee, C.F., \& Ostriker, J.P. 1977, \apj, 218, 148

\item Mihos, J.C., \& Hernquist, L. 1994, \apj, 437, 611

\item Mihos, J.C., \& Hernquist, L. 1996, \apj, 464, 641

\item Minter, A.H., \& Spangler, S.R. 1997, \apj, 485, 182

\item Myers, P.C. 1978, \apj, 225, 380

\item Navarro, J.F., \& White, S.D.M. 1993, \mnras, 265, 271

\item Norman, C.A., \& Ferrara, A. 1996, \apj, 467, 280

\item Novikov, Y. A. 1978, Izv. Atmos. Ocean. Phys., 14, 474  

\item Padoan, P., \& Nordlund, A. 1999, \apj, 526, 279

\item Passot, T., V\'azquez-Semadeni, E., \& Pouquet A. 1995, \apj,
455, 533

\item Passot, T. \& Pouquet, A. 1988, J. Comput. Phys. 75, 300

\item Reynolds, R.J. 1997, in Proc. of 156 WE-Heraeus-Seminar on 
The Physics of Galactic Halos, eds. H. Lesch et al. (Berlin:
Akademie Verlag), p. 57

\item Rosen, A., \& Bregman, J.N. 1995, \apj, 440, 634

\item Santangelo, P., Benzi, R. \& Legras, B. 1989, Phys. Fluids A, 1, 1027 

\item Scalo, J.M. 1987, in Interstellar processes, ed. D.J.Hollenbach
\& H.A.Thronson (Dordrecht: Reidel), 349

\item Scalo, J.M., \& Struck-Marcell, C. 1984, \apj, 276, 60

\item Silich, S.A. Franco, J., Palous, J., \& Tenorio-Tagle 1996,
\apj, 468, 722

\item Slavin, J.D., \& Cox, D.P. 1992, \apj, 392, 131

\item Slavin, J.D. 2000, in 
Astrophysical plasmas: codes, models, and observations, 
eds. J. Arthur, N. Brickhouse, \& J. Franco, RevMexAA 
(Serie de Conferencias), v. 9, 246

\item Slavin, J.D., Shull, J.M., \& Begelman, M.C. 1993, \apj, 407, 83

\item  Somerville, R.S., \& Primack, J.R. 1999, \mnras, 310, 1087

\item Stone, J.M., Ostriker, E.O., \& Gammie, C.F. 1998, \apj, 508, L99

\item Struck, C., \& Smith, D.C. 1999, \apj, 527, 673

\item Tammann, G.A., Loffler, W., \& Schroeder, A. 1994, ApJS, 92, 487 

\item Tomisaka, K. 1992, PASJ, 44, 177

\item Tomisaka, K. 1998, \mnras,  298, 797

\item van den Bosch, F.C. 2000, \apj, 530, 177

\item V\'azquez, E.C. \& Scalo, J.M. 1989, \apj, 343, 644

\item \VS, E. 1999, in Millimeter \& Submillimeter Astronomy:
   Chemistry and Physics in Molecular Clouds, 
   Proc. of the 1996 INAOE Summer School of Millimeter-Wave Astronomy,
   eds. W. F. Wall, A. Carrami\~nana, L. Carrasco, and P. F. Goldsmith
   (Dordrecht: Kluwer), 161

\item \VS, E., Gazol, A. \& Scalo, J. 2000, \apj, 540, 271    

\item V\'azquez-Semadeni, E., Passot, T., \& Pouquet A. 1995, \apj,
441, 702

\item V\'azquez-Semadeni, E., Passot, T., \& Pouquet A. 1996, \apj, 473, 881

\item \VS, Ostriker, E.C., Passot, T., Gammie, C.F.,
 \& Stone, J.M. 2000, in Protostars and Planets IV, eds. V.Mannings, 
A.Boss, S.Russell (Tucson: Univ.\ of Arizona Press), p. 3 

\item Wang, B., \& Silk, J. 1994, \apj, 427, 759

\item  White, S.D.M, \& Frenk, C.S. 1991, \apj, 379, 52

\end{description}

\clearpage

\vfill
\eject

{ \footnotesize
\begin{deluxetable}{cccccccccccc}
\tablecaption{Direct and indirect energy injection parameters}
\tablehead{
Run $\#$ & Resolution\tablenotemark{a} & $l_{\rm f}$/pc	&$\uf/$\kms
&$\epsdot$\tablenotemark{b}	&$\Es/10^{49}$erg\tablenotemark{c}
&$\langle \sfrOB \rangle$\tablenotemark{d}	&$f$\tablenotemark{e}	
}
\startdata
\\
6 (ISM)\tablenotemark{f}	&128	& 30.0	& $\sim7.0$  & ---    & ---    &$\sim14.0$   & --- 	
\\
12 (dec.)\tablenotemark{g}	&128	& NA	& NA	& NA	& NA	& NA	& NA 	
\\
17 (512 dec.)	&512	& NA	& NA	& NA	& NA	& NA	& NA 	
\\
30 (fiducial)	&128	& 66.5	& 7.02	& 1.99	& 2.84	& 0.71 	& .36
\\
33		&128	& 38.0	& 6.44	& 2.82	& 1.31	& 0.49 	& .10
\\
34		&128	& 66.5	& 3.21	& 0.38	& 0.54	& 0.41 	& .26
\\
35		&128	& 66.5	& 7.02	& 1.99	& 2.84	& 1.63 	& .55
\\
36		&128	& 66.5	& 3.21	& 0.38	& 0.54	& 1.02 	& .45
\\
37		&128	& 38.0	& 6.44	& 2.82	& 1.31	& 1.13 	& .18
\\
38		&128	& 38.0	& 3.21	& 0.48	& 0.23  & 0.61 	& .11
\\
39		&128	& 38.0	& 3.21	& 0.48	& 0.23  & 0.28 	& .06
\\
40		&128	& 133.	& 2.92	& 0.22	& 1.25	& 0.58 	& .67
\\
41		&128	& 133.	& 4.10	& 0.69	& 3.94	& 0.71 	& .68
\\
42		&128	& 66.5	& 7.02	& 1.99	& 2.84	& 0.10 	& .09
\\
43		&128	& 38.0	& 1.64	& 0.12	& 0.06  & 0.45 	& .09
\\
44 (512 fid.)	&512	& 66.5	& 5.86	& 1.63	& 2.33	& 0.73	& .35
\\
\enddata
\tablenotetext{a} {Number of grid points per dimension}
\tablenotetext{b} {In units of $10^{-3}$ erg s$^{-1}$ gr$^{-1}$}
\tablenotetext{c} {Total energy injected per stellar event}
\tablenotetext{d} {Time-averaged ``OB-star'' formation rate (in units
of $10^{-5}$ yr$^{-1}$ kpc$^{-2}$), from $t=40.8$ to $t=272$ Myr.}
\tablenotetext{e} {Measured filling factor of all forcing regions}
\tablenotetext{f} {``---'' denotes data not measured for this run}
\tablenotetext{g} {NA means not applicable}

\end{deluxetable}
}

\vfill
\eject

{ \footnotesize
\begin{deluxetable}{cccccccccccccccc}
\tablecaption{Global quantities measured in the runs}
\tablehead{
Run $\#$\tablenotemark{a} &	$\langle\Edotin\rangle$\tablenotemark{b}	&$\langle\Ek\rangle$\tablenotemark{c}	&$\langle\urms\rangle/$\kms	&$\langle\td\rangle/$Myr	& $\langle\tin\rangle$\tablenotemark{d}		&$\tin$(pred.)\tablenotemark{e}    &$\lambda_{\rm E}$/pc\tablenotemark{f}   &$t_{\rm cr}$\tablenotemark{g}		
}
\startdata
6 (ISM)\tablenotemark{h} & $\sim2.8$     & 1.80	& 6.8	& 18.0	& 20.4  & NA      & 245.\tablenotemark{i}	  & 34.7
\\
12 (dec.)	 & NA	& NA	& NA	& NA	& 18.0 	& NA    & ---      & ---
\\
17 (512 dec.)	 & NA	& NA	& NA	& NA	& 16.6 	& NA	& ---	  & ---
\\
30 (fiducial)	 & 2.36	& 1.20	& 5.5	& 16.5	& 16.3 	& 13.8  & 163.   & 27.4
\\
33		 & 0.80	& 0.32	& 3.0	& 13.0	& 12.8 	& 11.1  & 107.   & 34.4
\\
34	     	 & 0.33	& 0.25	& 2.6	& 25.0	& 24.3 	& 21.8  & 178.   & 66.1
\\
35		 & 5.43	& 2.39	& 7.6	& 15.0	& 14.2 	& 17.3  & 151.5   & 19.4
\\
36		 & 0.85	& 0.50	& 3.5	& 19.0	& 18.9 	& 23.3  & 151.5   & 45.0
\\
37		 & 1.56	& 0.59	& 4.0	& 12.0	& 12.1 	& 10.8  & 107.   & 26.2
\\
38		 & 0.15	& 0.06	& 1.3	& 13.0  & 13.0 	& 10.8  & 108.   & 84.5
\\
39		 & 0.07	& 0.03	& 0.9	& 13.0  & 13.2 	& 10.8  & 107.   & 115.
\\
40		 & 0.90	& 0.86	& 4.4	& 33.0	& 30.8 	& 48.4  & 244.5   & 55.1
\\
41		 & 1.84	& 1.69	& 6.2	& 31.0	& 29.6 	& 42.1  & 271.   & 46.4
\\
42		 & 0.40	& 0.33	& 3.0	& 28.0	& 26.4 	& 16.2  & 173.   & 50.3
\\
43		 & 0.03	& 0.01	& 0.6	& 14.6  & 14.0 	& 10.8  & 104.   & 178.
\\
44 (512 fid.)	 & 2.15	& 1.07	& 4.9	& 16.0	& 16.0  & 13.7  & 143.\tablenotemark{j}	& 26.6
\\
\\
\enddata
\tablenotetext{a} {All the measured quantities are averages over $ 40.8 \le t \le 
272$ (Myr) and correspond to a total area of 1 kpc$^2$}
\tablenotetext{b} {Energy injection rate (in units of $10^{36}$ erg/s)}
\tablenotetext{c} {Energy content (in units of erg)}
\tablenotetext{d} {Measured characteristic injection time scale (in Myr)}
\tablenotetext{e} {Predicted characteristic injection time scale according
to eq. (\ref{eq:tifin}) (in Myr)}
\tablenotetext{f} {Length scale corresponding to the energy 
spectrum centroid (see eq. [\ref{eq:centroid}]) measured at $t=136$
Myr (51.7 Myr for run 44)}
\tablenotetext{g} {Large-scale crossing time $t_{\rm cr}\equiv \lambda_{\rm
e}/\urms$ (in Myr)}
\tablenotetext{h} {``---'' and NA mean not measured and not applicable, 
respectively}
\tablenotetext{i} {Centroid calculated at $t=272$ Myr to allow for
a roughly stationary regime in this run}
\tablenotetext{j} {Centroid calculated at $t=68.0$ Myr (maximum
evolution time for this run)}
\end{deluxetable}
}

\end{document}